\documentclass[10pt]{amsart}
\usepackage{amsmath,amsfonts,amssymb,amsthm,amscd}
\newtheorem{thm}{Theorem}
\newtheorem{lem}[thm]{Lemma}
\newtheorem{defn}[thm]{Definition}
\newtheorem{prop}[thm]{Proposition}
\newtheorem{cor}[thm]{Corollary}
\newtheorem{example}[thm]{Example}
\newtheorem{conjecture}[thm]{Conjecture}

\def\res{\mathop{\rm res}\limits}

\begin{document}

\title[Weyl groups and Elliptic Solutions of the WDVV equations]{Weyl groups and Elliptic Solutions of the WDVV equations}

\author{Ian A. B. Strachan}


\date{\today}

\address{Department of Mathematics\\ University of Glasgow\\ Glasgow G12 8QQ\\ U.K.}

\email{i.strachan@maths.gla.ac.uk}

\keywords{Frobenius manifolds, WDVV equations, Jacobi groups, elliptic functions, elliptic polylogarithms}
\subjclass{11F55, 53B50, 53D45}

\begin{abstract}
A functional ansatz is developed which gives certain elliptic solutions of the Witten-Dijkgraaf-Verlinde-Verlinde (or WDVV)
equation. This is based on the elliptic trilogarithm function introduced by Beilinson and Levin. For this to be
a solution results in a number of purely algebraic conditions on the set of vectors that appear in the ansatz, this
providing an elliptic version of the idea, introduced by Veselov, of a $\vee$-system.

Rational and trigonometric limits are studied together with examples of elliptic $\vee$-systems based on various Weyl
groups. Jacobi group orbit spaces are studied: these carry the structure of a Frobenius manifold. The corresponding
\lq almost dual\rq~structure is shown, in the $A_N$ and $B_N$ and conjecturally for an arbitrary Weyl group, to
correspond to the elliptic solutions of the WDVV equations.

Transformation properties, under the Jacobi group, of the elliptic trilogarithm are derived together with
various functional identities which generalize the classical Frobenius-Stickelburger relations.
\end{abstract}

\maketitle

\tableofcontents

\section{Introduction}

One recurrent theme in the theory of integrable systems is the tower of
generalizations
\[
{\rm rational~}\longrightarrow{\rm trigonometric~}\longrightarrow{\rm elliptic}\,,
\]
the paradigm being provided by the Calogero-Moser system, where the original rational
interaction term may be generalized
\[
\frac{1}{z^2} \longrightarrow \frac{1}{\sin^2z} \longrightarrow \wp(z)
\]
whilst retaining integrability. A second recurrent theme is the appearance of root
systems, the paradigm being again provided by the Calogero-Moser system where the
interaction term
\[\displaystyle{\sum_{i\neq j} \frac{1}{(z_i-z_j)^2}}\]
can, on fixing the centre of mass,
be written as
\[
\displaystyle{\sum_{\alpha \in \mathcal{R}_{A_N}}
\frac{1}{(\alpha,{\bf z})^2}}\,,
\]
where the sum is taken over the roots $\mathcal{R}_{A_N}$ of the
$A_N$ Coxeter group \cite{OP}. The integrability of the system is preserved
if other root systems are used.

These two themes occur in many other integrable structures; $R$ matrices, quantum
groups, Dunkl operators, KZ-equations all admit (to a greater or lesser extent)
rational, trigonometric and elliptic versions and generalizations to arbitrary root
systems (see for example \cite{cher} and the references therein). In this paper elliptic solutions of the Witten-Dijkgraaf-Verlinde-Verlinde
(or WDVV) equations will be studied for arbitrary Weyl groups, these sitting at the right of
the following tower of generalizations:
\[
\begin{array}{ccccc}
\mathbb{C}^N/W & \longrightarrow & \mathbb{C}^{N+1}/{\widetilde{W}} & \longrightarrow & \Omega/J(\mathfrak{g})\,. \\
 &  &  &  &  \\
\left\{
\begin{array}{c}{\rm Coxeter~group}\\{\rm orbit~space} \end{array}
\right\} & \longrightarrow & \left\{
\begin{array}{c}{\rm Extended~affine~Weyl}\\{\rm orbit~space} \end{array}
\right\} & \longrightarrow & \left\{
\begin{array}{c}{\rm Jacobi~group}\\{\rm orbit~space} \end{array}
\right\} \\
\end{array}
\]
We begin by defining a Frobenius manifold.

\subsection{Frobenius Manifolds and almost-duality}

\begin{defn}
An algebra $(\mathcal{A},\circ,\eta,e)$ over $\mathbb{C}$ is a Frobenius algebra if:
\begin{itemize}
\item{} the algebra $\{\mathcal{A},\circ\}$ is commutative, associative with unity $e\,;$
\item{} the multiplication is compatible with a $\mathbb{C}$-valued bilinear, symmetric, nondegenerate
inner product
\[
\eta\,: \,\mathcal{A}\times \mathcal{A}\rightarrow\mathbb{C}
\]
in the sense that
\[
\eta(a \circ b,c) = \eta(a,b\circ c)
\]
for all $a,b,c\in\mathcal{A}\,.$
\end{itemize}
\end{defn}

\noindent With this structure one may define a Frobenius manifold
\cite{dubrovin1}:
\begin{defn} $(M,\circ,e,\eta,E)$ is a Frobenius manifold if each tangent
space $T_pM$ is a Frobenius algebra varying smoothly over $M$ with the
additional properties:
\begin{itemize}
\item{} the inner product is a flat metric on $M$ (the term \lq metric\rq~will denote a
complex-valued quadratic form on $M$).
\item{} $\nabla e=0$, where $\nabla$ is the Levi-Civita connection of the metric;
\item{} the tensor $(\nabla_W \circ)(X,Y,Z)$ is totally symmetric for all vectors
$W,X,Y,Z \in TM\,;$
\item{} the vector field $E$ (the Euler vector field) has the properties
\[
\nabla(\nabla E) = 0\]
and the corresponding one-parameter group of diffeomorphisms acts by
conformal transformations of the metric and by rescalings on the Frobenius
algebras $T_pM\,.$
\end{itemize}

\end{defn}
\noindent Since the metric $\eta$ is flat there exists a
distinguished coordinate system (defined up to linear
transformations) of so-called flat coordinates\footnote{This
labeling is for future notational convenience.}
$\{t^\alpha\,,\alpha=0\,,\ldots\,,N+1\}$ in which the components
of the metric are constant. From the various symmetry properties
of tensors $\circ$ and $\nabla\circ$ it then follows that there
exists a function $F$, the prepotential, such that in the flat
coordinate system,
\begin{eqnarray*}
c_{\alpha\beta\gamma} & = &
\eta\left(\frac{\partial~}{\partial t^\alpha} \circ \frac{\partial~}{\partial t^\beta},\frac{\partial~}{\partial t^\gamma}\right)\,,\\ & = &
\frac{\partial^3F}{\partial t^\alpha \partial t^\beta \partial t^\gamma}\,,
\end{eqnarray*}
and the associativity condition then implies that the pair $(F,\eta)$ satisfy the WDVV-equations
\[
\frac{\partial^3F}{\partial t^\alpha \partial t^\beta \partial t^\lambda}\eta^{\lambda\mu}
\frac{\partial^3F}{\partial t^\mu \partial t^\gamma \partial t^\delta}-
\frac{\partial^3F}{\partial t^\delta \partial t^\beta \partial t^\lambda}\eta^{\lambda\mu}
\frac{\partial^3F}{\partial t^\mu \partial t^\gamma \partial t^\alpha}=0\,,\quad
\]
where $\alpha\,,\beta\,,\gamma\,,\delta=0\,\ldots\,,N+1\,.$

Consider the vector field $E^{-1}$ defined by the condition
\[
E^{-1} \circ E = e\,.
\]
This is defined on $M^{\star}= M\backslash \Sigma\,,$ where $\Sigma$ is the discriminant submanifold
on which $E^{-1}$ is undefined. With this field one may define a new \lq dual\rq~multiplication
$\star: TM^\star \times TM^\star \rightarrow TM^\star$ by
\[
X \star Y = E^{-1} \circ X \circ Y\,, \qquad\qquad \forall\, X\,,Y \in TM^\star\,.
\]
This new multiplication is clearly commutative and associative, with the Euler vector field being the
unity field for the new multiplication.

Furthermore, this new multiplication is compatible with the intersection form $g$ on the Frobenius manifold,
i.e.
\[
g(X\star Y, Z) = g(X,Y\star Z)\,, \qquad\qquad \forall\, X\,,Y\,,Z \in TM^\star\,.
\]
Here $g$ is defined by the equation
\[
g(X,Y)=\eta(X\circ Y, E^{-1})\,, \qquad\qquad \forall\, X\,,Y \in TM^\star
\]
(and hence is well-defined on $M^\star\,$). Alternatively one may use the metric $\eta$ to extend the
original multiplication to the cotangent bundle and define
\[
g^{-1}(x,y) = \iota_E(x\circ y) \,, \qquad\qquad \forall\, x\,,y \in T^\star M^\star\,.
\]
The intersection form has the important property that it is flat, and hence there exists a distinguished
coordinate system $\{ {\bf p}\}$ in which the components of the intersection form are constant. It turns
out that there exists a dual prepotential $F^\star$ such that its third derivatives give the structure
functions $c^{\star}_{ijk}$ for the dual multiplication. More precisely \cite{dubrovin2}:

\begin{thm} Given a Frobenius manifold $M$, there exists a function $F^\star$ defined on $M^\star$
such that:
\begin{eqnarray*}
c^{\star}_{ijk} & = &
g\left( \frac{\partial~}{\partial p^i}\star \frac{\partial~}{\partial p^j}\,, \frac{\partial~}{\partial p^k}
\right) \,,\\
& = &\frac{\partial^3 F^\star}{\partial p^i \partial p^j \partial p^k}\,.
\end{eqnarray*}
Moreover, the pair $(F^\star,g)$ satisfy the WDVV-equations in the flat coordinates $\{ {\bf p} \}$ of the metric $g\,.$
\end{thm}

Thus given a specific Frobenius manifold one may construct a \lq dual\rq~solution
to the WDVV-equations by constructing the flat-coordinates of the intersection
form and using the above result to find the tensor $c^{\star}_{ijk}$ from which the
dual prepotential may be constructed.

\subsection{Examples}

The simplest class of Frobenius manifolds is given by the so-called Saito construction
on the space of orbits of a Coxeter group. Let $W$ be an irreducible Coxeter group acting on a real vector space
$V$ of dimension $N\,.$ The action extends to the complexified space $V \otimes \mathbb{C}\,.$ The
orbit space
\[
V\otimes\mathbb{C}\slash W \cong \mathbb{C}^{N}\slash W
\]
has a particularly nice structure, this following from Chevalley's theorem
on the ring of $W$-invariant polynomials:
\begin{thm}\label{chevalley}
There exists a set of $W$-invariant polynomial $s_i({\bf z})\,,i=1\,,\ldots\,,N$ such that
\[
\mathbb{C}[z_1\,,\ldots\,,z_N]^W \cong
\mathbb{C}[s_1\,,\ldots\,,s_N]\,.
\]
\end{thm}

On this
orbit space one may define a metric (a complex-valued quadratic
form) by taking the Lie-derivative of the $W$-invariant Euclidean
metric $g$ on $V\otimes\mathbb{C}$
\[
\eta^{-1} =\mathcal{L}_e g^{-1}
\]
where $e$ is a vector field constructed from the highest degree
invariant polynomial. It was proved by K. Saito that this metric
is non-degenerate and flat \cite{saito}. One therefore obtains a flat pencil of
metrics from which one may construct a polynomial solution -
polynomial in the flat coordinates of the metric $\eta$ - to the WDVV
equations.

The dual prepotential for this class of Frobenius manifolds is particularly
simple:
\begin{equation}
\label{dualrational}
F=\frac{1}{4} \sum_{\alpha \in \mathcal{R}_W} (\alpha,{\bf z})^2 \log (\alpha,{\bf z})^2\,
\end{equation}
where the sum is taken
over the roots of the Coxeter group $W$ \cite{Marsh, Marsh2, Martini}. However, the space of solutions of the same
functional form is far larger. Veselov \cite{Veselov} derived the algebraic conditions, known
as $\vee$-conditions, on the set of vectors $\mathfrak{U}$ that are required for
the prepotential
\[
F=\frac{1}{4} \sum_{\alpha \in \mathfrak{U}} (\alpha,{\bf z})^2 \log (\alpha,{\bf z})^2\,
\]
to satisfy the WDVV equations (we assume throughout this paper that if $\alpha \in \mathfrak{U}$ then $-\alpha \in \mathfrak{U}$ automatically).
What is required here is a
refinement of this idea, namely that of a complex Euclidean
$\vee$-system \cite{FV2}
\begin{defn}\label{bigvee}
Let $\mathfrak{h}$ be a complex vector space with non-degenerate
bilinear form $(\,,\,)$ and let $\mathfrak{U}$ be a collection of
vectors in $\mathfrak{h}\,.$ A complex Euclidean $\vee$-system
$\mathfrak{U}$ satisfies the following conditions:
\begin{itemize}
\item{} $\mathfrak{U}$ is well distributed, i.e.
$\sum_{\alpha\in\mathfrak{U}} h_\alpha (\alpha,{\bf u})
(\alpha,{\bf v}) = 2 h^\vee_\mathfrak{U} ({\bf u},{\bf v})$ for
some $\lambda\,;$ \item[] \item{} on any 2-dimensional plane $\Pi$
the set $\Pi\cap\mathfrak{U}$ is either well distributed or
reducible (i.e. the union of two non-empty orthogonal subsystems).
\end{itemize}
\end{defn}
\noindent Note the following:

\begin{itemize}

\item{} the constants $h_\alpha$ could be absorbed into the
$\alpha\,.$ In applications these constants will be both positive
and negative. Hence the requirement of a complex vector space.

\item{} the constant $h^\vee_\mathfrak{U}$ can be zero in certain
spaces.

\end{itemize}

\noindent One further comment has to be made in the case when
$h^\vee_\mathfrak{U}=0\,.$ We require here that the inverse metric
used in the WDVV equations is the non-degenerate bilinear form
$(\,,)$ on $\mathfrak{h}\,$ rather than one - possibly degenerate
- constructed from the sum of derivatives of $F$ as used in
\cite{FV}\,.

Trigonometric solutions were studied in \cite{DZ}, corresponding to extended affine Weyl
groups. As in the Coxeter case one has a Chevalley-type theorem and a well defined orbit
space on which one may define, following the Saito-construction, a flat metric and hence
a solution to the WDVV equations. It is to be expected, though a full proof for arbitrary Weyl groups is currently lacking,
that the corresponding dual
solutions will take the following functional form
\begin{equation}
\label{dualtrig}
F={\rm cubic~terms~} + \sum_{\alpha \in \mathcal{R}_W} h_{\alpha} Li_3\left(e^{i (\alpha,{\bf x})}\right)
\end{equation}
where $Li_3(x)$ is the trilogarithm and $h_\alpha$ are Weyl-invariant sets of constants. Solutions of the
WDVV equations of this type have been studied by a number of authors \cite{Marsh2,MH} but are only known to be
almost dual solutions to the extended affine Weyl Frobenius manifolds in certain special cases (e.g. $W=A_N^{(k)}\,$) \cite{thesis}.
Trigonometric
$\vee$-conditions, conditions on the vectors $\alpha$ that ensure that the prepotential
\[
F={\rm cubic~terms~} + \sum_{\alpha \in \mathfrak{U}} h_{\alpha} Li_3\left(e^{i (\alpha,{\bf x})}\right)
\]
satisfies the WDVV equations, have also been studied recently \cite{misha}.

Elliptic solutions were studied in \cite{B}, being defined on the Jacobi group orbit space
$\Omega/J(\mathfrak{g})$. Further details and definitions will be given in Section \ref{hurwitzdetails},
following \cite{B},\cite{EZ} and \cite{wirth}. The Jacobi group
$J(\mathfrak g)$ (where $\mathfrak g$ is a complex finite dimensional simple Lie algebra of rank $N$
with Weyl group $W$) acts on the space
\[
\Omega = \mathbb{C}\oplus {\mathfrak h} \oplus \mathbb{H}
\]
where $\mathfrak h$ is the complex Cartan subalgebra of $\mathfrak g$ and $\mathbb{H}$ is the
upper-half-plane, and this leads to the study of invariant functions - the Jacobi forms.
Analogous to the Coxeter case, the orbit space
\[
\Omega/J({\mathfrak g})
\]
is a manifold and carries the structure of a Frobenius manifold. In \cite{RS}\,, using the
Hurwitz space description (see Section \ref{HurwitzSection})
\[
\Omega/J(A_N) \cong H_{1,N+1}(N+1)\,,
\]
the dual prepotential was constructed.

\begin{thm}\cite{RS}
\label{RileyStrachan}
The intersection form on the space $\Omega/J(A_N)$ is given by the formula
\[
g=2 du\,d\tau - \left.\sum_{i=0}^N (dz^i)^2\right|_{\sum_{j=0}^N z^j=0}\,
\]
(where $u \in \mathbb{C}\,,{\bf z} \in \mathfrak{h}$ and $\tau \in \mathbb{H}\,$).
The dual prepotential is given by the formula\footnote{Note, ${\sum_j}^{'}$ includes the term $j=0\,.$}
\begin{eqnarray*}
F^\star(u\,,{\bf z}\,,\tau) & = &
\frac{1}{2} \tau u^2 -
\frac{1}{2} u \sum_{i=0}^N (z^i)^2\\
& &
+\frac{1}{2}{\sum_{i\neq j}}^{'} \frac{1}{(2 \pi i)^3}
\left\{\mathcal{L}i_3(e^{2i(z^i-z^j)},e^{2\pi i \tau}) - \mathcal{L}i_3(1,e^{2\pi i \tau})\right\}
\\
&&-(N+1){\sum_j}^{'} \frac{1}{(2 \pi i)^3}
\left\{\mathcal{L}i_3(e^{2iz^j},e^{2\pi i \tau}) - \mathcal{L}i_3(1,e^{2\pi i \tau})\right\}\,.
\end{eqnarray*}
where this function is evaluated on the plane
$\sum_{j=0}^N z^j=0\,.$
\end{thm}
\noindent The precise definitions of the various terms in these formulae will be given below, but for now we note that
this dual prepotential
is given in terms of the
elliptic trilogarithm $\mathcal{L}i_3(z,q)$ introduced by Beilinson and Levin \cite{BL,Levin}. This function has
appeared already in the theory of Frobenius manifolds in the enumeration of curves {\cite{Kawai}.

This result is curious - as well as the $A_N$ root vectors appearing in the solution certain
extra vectors (in fact weight vectors) appear: these do not appear in the corresponding rational
and trigonometric solutions. This work raised a number of questions:
\begin{itemize}
\item{} Is there a direct verification that the function that appears in Theorem \ref{RileyStrachan}
satisfies the WDVV equations? Recall that its construction was via a Hurwitz space construction in terms of
certain holomorphic maps between the complex torus and the Riemann sphere.
\item{} What is the origin of the \lq extra\rq~vectors in the solution?
\item{} Can one construct solutions for other Weyl groups?
\end{itemize}
The purpose of this paper is to study solutions of the WDVV equations which take the
functional form
\[
F(u,{\bf z}, \tau) = \frac{1}{2} u^2 \tau - \frac{1}{2} u ({\bf z},{\bf z}) + \sum_{\alpha\in\mathfrak{U}} h_{\alpha} f\left(\, {\bf z}_\alpha, \tau\right)\,,
\]
where
\[
f(z,\tau)=\frac{1}{(2 \pi i)^3} \left\{\mathcal{L}i_3(e^{2 \pi i z},e^{2 \pi i\tau})-\mathcal{L}i_3(1,e^{2 \pi i \tau})\right\}\,,
\]
deriving a set of elliptic $\vee$-conditions on the \lq roots\rq~contained in the set $\mathfrak{U}\,.$ Thus the
above questions can all be answered affirmatively. This leaves the following question:
\begin{itemize}
\item{} For which elliptic $\vee$-systems is the solution the almost-dual solution to the
Jacobi group orbit space $\Omega/J({\mathfrak g})\,?$
\end{itemize}
This question has been answered already in the $A_N$ case \cite{RS} and in this paper
we extend the results to the $B_N$ case. For other Weyl groups it remains an open problem.


\section{The Elliptic Polylogarithm and its Properties}

The functional form of the above prepotential uses the elliptic polylogarithm. In this
section this is defined and its transformation properties under shifts and
modular transformations are studied. Before this we define various special
functions and the notation that will be used throughout the rest of this paper.

\subsection{Notation} There are, unfortunately, many different definitions and
normalizations for elliptic, number-theoretic and other special functions. Here
we list the definitions used in this paper. Let $q=e^{2\pi i \tau}\,,$ where $\tau \in\mathbb{H}\,.$

\begin{itemize}

\item{} $\vartheta_1$-function:
\[
\vartheta_1(z|\tau) = -i \left(e^{\pi i z}-e^{-\pi i z}\right) q^{\frac{1}{8}} \prod_{n=1}^\infty
(1-q^n) \left(1-q^n e^{2 \pi i z}\right)\left(1-q^n e^{-2\pi i z}\right)\,.
\]
The fundamental lattice is generated by $z\mapsto z+1\,,z\mapsto z+\tau\,,$ and the function
itself satisfies the complex heat equation
\[
\frac{\partial^2\vartheta_1}{\partial z^2} =4 \pi i \frac{\partial\vartheta_1}{\partial \tau}\,.
\]

\item{} Bernoulli numbers and Bernoulli polynomials:

\[
\frac{x}{e^x-1} = \sum_{n=0}^\infty B_n \frac{x^n}{n!}\,,\qquad\qquad
B_n(z)=\sum_{k=0}^n \left( \begin{array}{c} n \\ k \end{array} \right) B_k z^{n-k}\,.
\]

\item{}Eisenstein series:

\[
E_k(\tau) = 1 - \frac{2k}{B_k} \sum_{n=1}^\infty \sigma_{k-1}(n)\, q^n\,,\qquad k \in 2\mathbb{N}
\]
where $\sigma_k(n)=\sum_{d|n} d^k\,.$

\item{}Dedekind $\eta$-function:

\[
\eta(\tau) = q^{\frac{1}{24}} \prod_{n=1}^\infty (1-q^n)\,.
\]

\item{} Polylogarithm function:
\[
Li_N(z) = \sum_{n=1}^\infty \frac{z^n}{n^N}\,, \qquad |z|<1\,.
\]

\end{itemize}
\noindent Note that $\vartheta_1\,,E_2$ and $\eta$ are related:
\[
\frac{ \eta^\prime(\tau)}{\eta(\tau)} = \frac{2 \pi i}{24}\, E_2(\tau) =
\frac{1}{12 \pi i} \frac{\vartheta_1^{\prime\prime\prime}(0,\tau)}{\vartheta_1^\prime(0,\tau)}\,.
\]

\noindent These have the following properties under inversion of the independent variable:

\begin{eqnarray*}
\tau^{-n} E_n\left(-\frac{1}{\tau}\right) & = & E_n(\tau) \,, \qquad n\geq 4\,; \\
\tau^{-2} E_2\left(-\frac{1}{\tau}\right) & = & E_2(\tau) + \frac{12}{2 \pi i \tau}\,;\\
\eta\left(-\frac{1}{\tau}\right) & = & \sqrt{\frac{\tau}{i}}\, \eta(\tau)
\end{eqnarray*}
where in the last formula the square-root is taken to have non-negative real part. The
polylogarithm has the inversion property (for $n\in\mathbb{N}$ - other more complicated
versions hold for other values):

\begin{equation}
Li_n\left( e^{2 \pi i z}\right) +(-1)^{n} Li_n\left(e^{-2\pi i z}\right) = - \frac{(2 \pi i)^n}{n!} B_n(z)\,.
\label{polyloginversion}
\end{equation}
This inversion formula holds: if $\Im(z) \geq 0$ for $0 \leq \Re(z) < 1\,,$ and if $\Im(z) < 0$ for $0 < \Re(z) \leq 1\,.$
This, and other similar formulae, may be used to analytically continue the function outside the unit disc to a multi-valued
holomorphic function on $\mathbb{C}\backslash \{0,1\}\,.$ For a discussion of the monodromy
of the polylogarithm function see \cite{Ram}. This multivaluedness will also occur in the
solution of the WDVV equations. However this multivaluedness occurs in the quadratic terms only
and hence any physical quantities are single-valued.

\subsection{The elliptic polylogarithm}

An \lq obvious\rq~elliptic generalization of the polylogarithm function is
\[
\mathcal{L}i_r(\zeta,q) = \sum_{n=-\infty}^\infty Li_r(q^n \zeta)\,.
\]
However this series diverges, but by using the inversion formula
(\ref{polyloginversion}) and $\zeta$-function regularization one can
arrive at the following definition of the elliptic polylogarithm
function \cite{BL,Levin}:

\[
\mathcal{L}i_r(\zeta,q) = \sum_{n=0}^\infty Li_r(q^n \zeta) +
\sum_{n=1}^\infty Li_r(q^n \zeta^{-1}) -
\chi_r(\zeta,q)\,,\qquad\quad r~{\rm odd}\,,
\]
where
\[
\chi_r(\zeta,q)=\sum_{j=0}^r \frac{B_{j+1}}{(r-j)!(j+1)!}
(\log\zeta)^{(r-j)} (\log q)^j\,.
\]
A real-valued version of this function had previously been studied by Zagier \cite{Z}.
With this the function $f$ may be defined.

\begin{defn}
\label{defnf}
The function $f(z,\tau)\,,$ where $z\in\mathbb{C}\,,\tau\in\mathbb{H}\,,$ is defined
to be:
\[
f(z,\tau)=\frac{1}{(2 \pi i)^3} \left\{\mathcal{L}i_3(e^{2 \pi i z},q)-\mathcal{L}i_3(1,q)\right\}\,.
\]
\end{defn}
\noindent It follows from the definitions that
\begin{equation}
\label{e4int}
\left( \frac{d~}{d\tau}\right)^3 \frac{1}{(2 \pi i)^3} \mathcal{L}i_3(1,q)= \frac{1}{120}  E_4(\tau)
\end{equation}
and
\begin{equation}
\label{fthetaeta}
\left( \frac{\partial~}{\partial z}\right)^2 f(z,\tau)  =  -\frac{1}{2 \pi i}  \log\left\{ \frac{\vartheta_1(z,\tau)}{\eta(\tau)}\right\}\,.
\end{equation}
Thus the elliptic-trilogarithm may be thought of as a classical function (or, at least, a
neoclassical function) as it may be obtained from classical functions via nested integration
and other standard procedures. It does, however, provide a systematic way to deal with the
arbitrary functions that would appear this way. The notation $F \simeq G$ will be used if the functions
$F$ and $G$ differ by a quadratic function in the variables $\{u,{\bf z},\tau\}\,$ (recall that any prepotential
satisfying the WDVV equations is only defined up to quadratic terms in the flat-coordintes).

\begin{prop}\label{trans} The function $f$ has the following transformation properties:
\begin{eqnarray*}
f(z+1,\tau) &\simeq & f(z,\tau)\,;\\
f(z,\tau+1) & \simeq & f(z,\tau)\,;\\
f(z+\tau,\tau) & \simeq & f(z,\tau) +  \left\{ \frac{1}{6} z^3 + \frac{1}{4} z^2 \tau + \frac{1}{6} z \tau^2 + \frac{1}{24} \tau^3\right\}\,;\\
f(-z,\tau) & \simeq & f(z,\tau)\,.
\end{eqnarray*}
The function also has the alternative expansions:
\begin{equation}
\begin{array}{rcl}
\label{fseries1}
f(z,\tau) & \simeq & \displaystyle{-\frac{1}{(2 \pi i)} \left\{ \frac{1}{2} z^2 \log z + z^2 \log \eta(\tau) \right\}} \\
&&\\
&&\displaystyle{+\frac{1}{(2\pi i)^3} \sum_{n=1}^\infty \frac{ (-1)^n E_{2n}(\tau) B_{2n} }{(2n+2)! (2n)}(2 \pi z)^{2n+2}}
\end{array}
\end{equation}
and
\begin{equation}
\begin{array}{rcl}
\label{fseries2}
f(z,\tau) & \simeq & \displaystyle{\frac{1}{(2\pi i)^3} Li_3\left(e^{2 \pi i z}\right)+\frac{1}{12} z^3 - \frac{1}{24} z^2 \tau }\\
&&\\
&&\displaystyle{-\frac{4}{(2\pi i)^3} \sum_{r=1}^\infty \left\{
\frac{q^r}{(1-q^r)} \right\} \frac{\sin^2 (\pi r z)}{r^3}}
\end{array}
\end{equation}
Furthermore,
\[
f\left(\frac{z}{\tau}, -\frac{1}{\tau}\right) \simeq \frac{1}{\tau^2} f(z,\tau) - \frac{1}{\tau^3} \frac{z^4}{4!}\,.
\]
\end{prop}

{\sl Proof}
The first two relations follow from the definition. The third and fourth use the
inversion formula for polylogarithms (\ref{polyloginversion}).

The proof of (\ref{fseries1}) and (\ref{fseries2}) just involves some careful resumming. Consider the first two terms in the definition
of $f\,:$
\begin{eqnarray*}
\sum_{n=0}^\infty Li_3(q^n e^{2 \pi i z}) + \sum_{n=1}^\infty Li_3(q^n e^{-2 \pi i z})  & = & Li_3(e^{2 \pi i z})+\\
& & 2 \sum_{s=0}^\infty
\frac{(-1)^s}{(2s)!} \left\{ \sum_{n,r=1}^\infty q^{nr} r^{2s-3} \right\} (2 \pi z)^{2s}\,.
\end{eqnarray*}
From this series (\ref{fseries2}) follows immediately. To obtain (\ref{fseries1}) one rearranges the terms.
The $s=0$ term cancels in the final expression and the remaining terms may be re-expressed in terms of
Eisenstein series (for $s>1$) or the Dedekind function (for $s=1$). Finally, using the result
\begin{eqnarray*}
\frac{1}{(2 \pi i)^3}\frac{d^3~}{dz^3} Li_3\left(e^{2 \pi i z}\right) & = & -\frac{1}{2} \left[1+\coth(\pi i z)\right]\,, \\
& = &-\left[ \frac{1}{2} +\frac{1}{(2 \pi z)} +  \sum_{n=1}^\infty \frac{B_{2n}}{(2n)!} (2 \pi i z)^{2n-1}\right]
\end{eqnarray*}
one may obtain a series for $Li_3\left(e^{2 \pi i z}\right)\,.$ Putting all these parts together
gives the series (\ref{fseries1}).
$~~\Box$

\begin{thm}
\label{littletheorem}
The function
\[
h(z,\tau)= f(2z,\tau)- 4 f(z,\tau)
\]
satisfies the partial differential equation
\[
h^{(3,0)} (z,\tau)\,h^{(1,2)} (z,\tau) - \left[ h^{(2,1)} (z,\tau) \right]^2 + 4  h^{(0,3)} (z,\tau) =0
\]
where
\[
h^{(m,n)}(z,\tau) = \frac{\partial^{n+m}h}{\partial z^m \partial \tau^n}\,.
\]
\end{thm}

{\sl Proof} Let
\[
\Delta(z,\tau)=h^{(3,0)} (z,\tau)\,h^{(1,2)} (z,\tau) - \left[ h^{(2,1)} (z,\tau) \right]^2 + 4 h^{(0,3)} (z,\tau)\,.
\]
Using the transformation properties in Proposition \ref{trans} one may derive the transformation
properties of the derivatives and hence for the combination $\Delta\,.$ While the individual terms have quite complicated
transformation properties, those for $\Delta$ are very simple:
\begin{eqnarray*}
\Delta(z+1,\tau) & = & \Delta(z,\tau)\,,\\ &&\\
\Delta(z+\tau,\tau) & = & \Delta(z,\tau)\,,\\ &&\\
\Delta\left(\frac{z}{\tau},-\frac{1}{\tau}\right) & = & \tau^4 \Delta(z,\tau)\,.
\end{eqnarray*}
The first two of these equations imply that the function $\Delta$ is doubly periodic. From the series expansion
in Proposition \ref{trans} it follows that $\Delta$ has no poles: the only term which has a pole is
$f^{(3,0)}$ and this cancels with the zero in $f^{(1,2)}\,.$ Thus $\Delta$ is doubly periodic with no poles and hence
must be a function of $\tau$-alone (i.e. a theta-constant). The remaining transformation property implies that
$\Delta$ is a modular function of degree 4. The $q$-series representation of the function $f$ in
Proposition \ref{trans} implies that $\Delta$ is actually a cusp-form. But the space of cusp-forms of degree 4 is
empty and hence $\Delta=0\,.$
$~~\Box$

\begin{cor}\label{A1solutionA} The pair
\begin{eqnarray*}
F(u,z,\tau) & = & \frac{1}{2} u^2 \tau - u z^2 + h(z,\tau)\,,\\
& = & \frac{1}{2} u^2 \tau - u z^2 + f(2z,\tau) -4 f(z,\tau)
\end{eqnarray*}
and
\[
g= 2 du \, d \tau - 2 dz^2
\]
satisfy the WDVV equations.
\end{cor}

{\sl Proof}
The WDVV equations for the above prepotential reduce to the single equation
$\Delta=0$ so the result follows immediately from the above Theorem.
$~~\Box$

Using the same methods it is straightforward to show that the function
\begin{equation}
\label{A1solutionB}
{\tilde h}(z,\tau)=\frac{3}{4(2 \pi i)^3} \left[\mathcal{L}i_3(e^{2 \pi i z},q)+\frac{3}{2} \mathcal{L}i_3(1,q)\right]
\end{equation}
also satisfies the equation $\Delta=0\,.$ It is, however, the function $h$ which defines the dual
prepotential to the $A_1$-Jacobi group orbit space - see Theorem \ref{RileyStrachan}.

\section{Transformation properties of the WDVV equations}

Recall that we seek a solution of the WDVV of the form
\begin{equation}
\label{defF}
F(u,{\bf z}, \tau) = \frac{1}{2} u^2 \tau - \frac{1}{2} u ({\bf z},{\bf z}) +
\sum_{\alpha\in\mathfrak{U}} h_{\alpha} f\left(\, {\bf z}_\alpha, \tau\right)
\end{equation}
with $f(z,\tau)$ being given by Definition \ref{defnf}.
Sometimes the notation ${\bf z}_\alpha$ will be used to denote
$({\bf z},\alpha)\,,$ especially for terms involving the function
$f\,.$ Thus $f(({\bf z},\alpha),\tau)$ will be written
$f({\bf z}_\alpha,\tau)$ or even $f({\bf z}_\alpha)\,.$ The coordinates $\{
t^\alpha\,,\alpha=0\,,1\,,\ldots N\,,N+1\}$ are defined to be
\begin{eqnarray*}
t^0 & = & u \,,\\
t^i & = & z^i \,, \quad i=1\,,\ldots N\,, \\
t^{N+1} & = & \tau\,.
\end{eqnarray*}
Latin indices will range from $1$ to $N$ and Greek from $0$ to $N+1\,$ so the dimension of the manifold is
$N+2\,,$ with $N\geq 1\,.$ In addition
$ u \in\mathbb{C}\,, {\bf z} \in \mathfrak{h}\cong \mathbb{C}^N\,,\tau \in \mathbb{H}\,,$
so $(u,{\bf z},\tau) \in \Omega\,.$
Later, $\mathfrak{h}$ will be the complex Cartan subalgebra of a simple Lie algebra $\mathfrak{g}$
of rank $N$ with Weyl group $W\,, $ but for now it may be thought of a just $\mathbb{C}^N\,.$
Also, $(\,,\,)$ denotes the standard Euclidean inner product on $\mathfrak{h}\,.$
It follows from the functional dependence on $t^0=u$ that $\partial_u$ is the unity vector
field and hence the metric on $\Omega$ is
\[
g = d \tau\,du + du \, d\tau  - (d{\bf z},d{\bf z})\,.
\]

One of the main ideas of this paper is to extend Theorem \ref{littletheorem} to higher dimension, using
doubly-periodicity and modular arguments to prove that the WDVV equations are
satisfied. To begin we require a detailed analysis of the WDVV equations themselves.

\subsection{Analysis of the WDVV equations}

The WDVV equations are the conditions for a commutative algebra to be associative. Thus
they may be written in terms of the vanishing of the associator
\[
\Delta[X,Y,Z]=(X\circ Y)\circ Z - X \circ (Y \circ Z)\,.
\]
Since in the case being considered we have a unity element these simplify further: if any of the vector field
is equal to the unity field then $\Delta$ vanishes identically. Since the vector field $\partial_\tau\in T\mathbb{H}$
is special (for example, it behaves differently to the other variables under modularity transformation), we decompose these equation further, taking the inner product with arbitrary vector fields to obtain scalar-valued equations.

\begin{prop}\label{wdvvbreakdown} The WDVV equations for a multiplication with unity field are equivalent to
the vanishing of the following functions:
\[
\begin{array}{ccl}
\Delta^{(1)}({\bf u},{\bf v}) & = & g(\partial_\tau \circ \partial_\tau, {\bf u} \circ {\bf v}) - g(\partial_\tau \circ {\bf u}, \partial_\tau \circ {\bf v}) \,, \\ &&\\
\Delta^{(2)}({\bf u},{\bf v},{\bf w}) & = & g(\partial_\tau \circ {\bf u}, {\bf v} \circ {\bf w}) - g(\partial_\tau \circ {\bf w}, {\bf u} \circ {\bf v})\,,\\ &&\\
\Delta^{(3)}({\bf u},{\bf v},{\bf w},{\bf x}) & = & g({\bf u} \circ {\bf v},{\bf w}\circ {\bf x}) - g({\bf u} \circ {\bf x}, {\bf v} \circ {\bf w})\,
\end{array}
\]
for all ${\bf u}\,,{\bf v}\,,{\bf w}\,,{\bf x}\in T\mathfrak{h}\,.$
\end{prop}
\noindent In terms of coordinate vector fields these conditions are:
\begin{eqnarray*}
\Delta^{(1)}_{ij} &=&g_{ij}\, c_{\tau\tau\tau} + g^{pq} \left\{ c_{\tau\tau p} \,c_{ijq} - c_{\tau ip} \,c_{\tau jq}\right\}\,, \\
\Delta^{(2)}_{ijk} & = & \left\{ g_{jk}\, c_{\tau\tau i} - g_{ij} \,c_{\tau\tau k} \right\} +
g^{pq} \left\{ c_{\tau i p}\, c_{jkq} - c_{\tau kp}\, c_{ijq}\right\}\,, \\
\Delta^{(3)}_{ijrs} & = & \left\{ g_{ij} \, c_{\tau rs} + g_{rs} \, c_{\tau ij}-g_{is} \, c_{\tau rj}-g_{rj} \, c_{\tau is}\right\} +
g^{pq} \left\{ c_{ijp} \,c_{rsq} - c_{isp} \,c_{rjq}\right\}\,
\end{eqnarray*}
where $g_{ij}=-(\partial_i,\partial_j)\,.$
The function $\Delta$ in theorem (\ref{littletheorem}) is, since ${\rm dim}_\mathbb{C}\mathfrak{h}=1$, proportional to
$\Delta^{(1)}({\bf x},{\bf x})\,.$

\subsection{Modular transformations of the structure functions}

\begin{lem}\label{modlemma}
Let
\begin{equation}
\label{modtrans}
\begin{array}{rcl}
{\hat u} & = & \displaystyle{u - \frac{({\bf z},{\bf z})}{2 \tau}}\,,\\
{\hat {\bf z}} & = & \displaystyle{\frac{{\bf z}}{\tau}}\,,\\
{\hat\tau} & = & \displaystyle{-\frac{1}{\tau}}\,.
\end{array}
\end{equation}
Then
\[
F({\hat u},{\hat{\bf z}},{\hat \tau})=\frac{1}{\tau^2} \left\{ F(u,{\bf z},\tau) - \frac{1}{2} u
\left( 2 u  \tau -  ({\bf z},{\bf z})\right)\right\}
\]
if and only if
\begin{equation}
\label{conditionA}
\sum_{\alpha \in\mathfrak{U}} h_{\alpha} (\alpha,{\bf z})^4 = 3 ({\bf z},{\bf z})^2\,.
\end{equation}
\end{lem}

{\sl Proof}
This follows immediately from Proposition \ref{trans}.
$~~\Box$

The origin of the transformation (\ref{modtrans}) comes from the study of symmetries of the
WDVV equations (see \cite{dubrovin1} Appendix B). A symmetry is a transformation
\begin{eqnarray*}
t^\alpha & \mapsto & {\hat{t}^\alpha}\,, \\
g_{\alpha\beta} & \mapsto & {\hat g}_{\alpha\beta}\,,\\
F & \mapsto &{\hat F}
\end{eqnarray*}
that acts on the solution space of the WDVV equations. In particular, (\ref{modtrans}) is just
the transformation, denoted $I$ in \cite{dubrovin1},
\begin{eqnarray*}
{\hat t}^0 & = & \frac{1}{2} \frac{t_\sigma t^\sigma}{t^{N+1}}\,,\\
{\hat t}^i & = & \frac{t^i}{t^{N+1}}\,, \quad i = 1\,,\ldots\,,N\,,\\
{\hat t}^{N+1} & = & - \frac{1}{t^{N+1}}\,,\\
&&\\
{\hat g}_{\alpha\beta} & = & {g}_{\alpha\beta}\,,\\
&&\\
{\hat F}({\hat{\bf t}})&  = &\left(t^{N+1}\right)^{-2} \left[ F({\bf t}) - \frac{1}{2} t^0 (t_\sigma t^\sigma) \right]
\end{eqnarray*}
which induces a symmetry of the WDVV equations. Up to a simple equivalence, $I^2 = \mathbb{I}\,.$
It follows from this and Lemma \ref{modlemma} that we are at the fixed point of this
involution and that, rather than telling one how to construct a new solution from a seed
solution, it gives the transformation property of the various functions under the modular
transformations. A simple modification of Lemma B.1 \cite{dubrovin1} immediately gives:

\begin{prop}\label{modularprop}

Suppose $F$ is given by equation (\ref{defF}) where condition (\ref{conditionA}) is assumed to hold. Let $c_{\alpha\beta\gamma}=\partial_\alpha\partial_\beta\partial_\gamma F({\bf t})\,.$
Then
\[
c_{\alpha\beta\gamma}({\bf z}, \tau+1)  =  c_{\alpha\beta\gamma}({\bf z}, \tau)
\]
and
\begin{eqnarray*}
c_{ijk}\left( \frac{{\bf z}}{\tau},-\frac{1}{\tau} \right) & = & \tau \,c_{ijk}({\bf z},\tau) - g_{ij} \,z_k - g_{jk}\, z_i - g_{ki}\, z_j \,, \\
c_{\tau ij}\left( \frac{{\bf z}}{\tau},-\frac{1}{\tau} \right) & = & \tau\, c_{ij\alpha}({\bf z},\tau)\,t^\alpha - \frac{1}{2} g_{ij}\,
(t_\sigma t^\sigma) - z_i z_j\,, \\
c_{\tau\tau i}\left( \frac{{\bf z}}{\tau},-\frac{1}{\tau} \right) & = & \tau\, c_{i\alpha\beta}({\bf z},\tau)t^\alpha t^\beta-z_i \,(t_\sigma t^\sigma)\,,\\
c_{\tau\tau\tau}\left( \frac{{\bf z}}{\tau},-\frac{1}{\tau} \right) & = & \tau \,c_{\alpha\beta\gamma}({\bf z},\tau) t^\alpha t^\beta t^\gamma - \frac{3}{4} (t_\sigma t^\sigma)^2\,.
\end{eqnarray*}
With these, the transformation properties of the functions $\Delta^{(i)}$ are
\[
\Delta^{(i)}({\bf z},\tau+1) = \Delta^{(i)}( {\bf z}, \tau ) \,, \qquad \qquad i=1\,,2\,,3
\]
and
\begin{eqnarray*}
\Delta^{(3)}_{ijrs} \left( \frac{{\bf z}}{\tau},-\frac{1}{\tau} \right) & = & \tau^2\, \Delta^{(3)}_{ijrs} ({\bf z},\tau)\,,\\
\Delta^{(2)}_{ijk} \left( \frac{{\bf z}}{\tau},-\frac{1}{\tau} \right) & = & \tau^3 \,\Delta^{(2)}_{ijk} ({\bf z},\tau) + \tau^2 z^r \Delta^{(3)}_{irkj}({\bf z},\tau)\,,\\
\Delta^{(1)}_{ij} \left( \frac{{\bf z}}{\tau},-\frac{1}{\tau} \right) & = & \tau^4 \,\Delta^{(1)}_{ij} ({\bf z},\tau) -\tau^3 \,z^r\left\{
\Delta^{(2)}_{ijr}({\bf z},\tau)+\Delta^{(2)}_{jir}({\bf z},\tau)\right\}  \\
& & +\tau^2 \,z^a z^b \Delta^{(3)}_{abij}({\bf z},\tau) \,.
\end{eqnarray*}
\end{prop}
{\sl Proof}
The proof is straightforward and uses the transformation properties of $f$ derived in Proposition \ref{trans}.
$~~\Box$

\noindent It is important to
note that these $\Delta^{(i)}$ are powers series, not Laurent series, in the $q$-variable. Again, this follows from the $q$-expansion of
$f$ given in proposition \ref{trans}.

\subsection{Periodicity properties of the structure functions}

We assume that there exists a vector ${\bf p} \in \mathfrak{h}$ such that $(\alpha,{\bf p}) \in \mathbb{Z}$ for all
$\alpha \in \mathfrak{U}\,.$ Later we will require the existence of a full $N$-dimensional lattice (the \lq weight lattice\rq~
associated to the \lq roots\rq~in $\mathfrak{U}$), but for now we just require a single such vector. From Proposition \ref{trans} it follows
that
\[
f\left( (\alpha,{\bf z}+{\bf p}), \tau \right)\simeq f\left( (\alpha,{\bf z}), \tau \right)
\]
and hence $F(u,{\bf z}+{\bf p},\tau) \simeq F(u,{\bf z},\tau)\,.$ Thus
\[
\Delta^{(i)}({\bf z}+{\bf p},\tau) = \Delta^{(i)}({\bf z},\tau)\,,\quad i=1\,,2\,,3\,.
\]
The calculation of the transformations under shifts ${\bf z} \mapsto {\bf z}+{\bf p}\tau$ requires more care.
\begin{prop}
Assume that the following conditions hold:
\[
\sum_{\alpha \in\mathfrak{U}} h_{\alpha} (\alpha,{\bf z})^4 = 3({\bf z},{\bf z})^2\,,
\]
and $(\alpha,{\bf p}) \in\mathbb{Z}$ for all
${\alpha}\in\mathfrak{U}\,.$ Then
\begin{eqnarray*}
h({\bf z}+{\bf p}\tau,\tau) &\simeq& h({\bf z},\tau)+\\
&&\\
&&\frac{1}{8}
\left\{
\begin{array}{c}
4({\bf p},{\bf z})({\bf z},{\bf z})+
\tau\left[ 4 ({\bf p},{\bf z})^2+2({\bf p},{\bf p})({\bf z},{\bf z})\right]\\
\\+
4 \tau^2 ({\bf p},{\bf z})({\bf p},{\bf p})+
\tau^3 ({\bf p},{\bf p})^2
\end{array}
\right\}
\end{eqnarray*}
where
\[
h({\bf z},\tau) = \sum_{\alpha\in\mathfrak{U}} h_{\alpha} f( {\bf z}_\alpha,\tau)\,.
\]
\end{prop}
{\sl Proof}
From Proposition \ref{trans} it follows by induction, for $n \in \mathbb{Z}\,,$ that
\[
f(z+ n \tau, \tau) \simeq f(z,\tau) + \frac{1}{24} \left( 4 n z^3 + 6 n^2 \tau z^2 + 4 n^3 \tau^2 z + n^4 \tau^3 \right)
\]
and since, by assumption, $(\alpha,{\bf p}) \in\mathbb{Z}\,,$ it follows immediately that
\begin{eqnarray*}
f\left( (\alpha,{\bf z}) +(\alpha,{\bf p}) \tau,\tau\right) & \simeq & f\left( (\alpha,{\bf z}),\tau \right) +\\
&&\frac{1}{24}
\left\{
\begin{array}{c}
4(\alpha,{\bf p})(\alpha,{\bf z})^3 + 6 \tau (\alpha,{\bf p})^2(\alpha,{\bf z})^2+\\ \\
4\tau^2 (\alpha,{\bf p})^3(\alpha,{\bf z}) + \tau^3(\alpha,{\bf p})^4
\end{array}
\right\}\,.
\end{eqnarray*}
Hence, on summing over $\alpha\,,$
\begin{eqnarray*}
h({\bf z} + {\bf p} \tau,\tau) & \simeq & h\left( {\bf z},\tau \right) +\\
&&\frac{1}{24}
\left\{
\begin{array}{c}
4\sum_\alpha h_\alpha (\alpha,{\bf p})(\alpha,{\bf z})^3 + 6 \tau \sum_\alpha h_\alpha (\alpha,{\bf p})^2(\alpha,{\bf z})^2+\\ \\
4\tau^2 \sum_\alpha h_\alpha (\alpha,{\bf p})^3(\alpha,{\bf z}) + \tau^3\sum_\alpha h_\alpha (\alpha,{\bf p})^4
\end{array}
\right\}\,.
\end{eqnarray*}
Hence, using the first condition (and its polarized version), the result follows.
$~~\Box$

With this the following Proposition may be proved: the first part is immediate and the second part follows from routine but tedious
calculations.
\begin{prop}\label{periodprop}
Under the conditions of the above proposition, the structure functions have the following transformation properties:
\begin{eqnarray*}
c_{ijk}({\bf z}+{\bf p}\tau,\tau) & = & c_{ijk}({\bf z},\tau)+ p_i \,g_{jk}+p_j \, g_{ki} +p_k\,g_{ij}\,, \\
c_{\tau ij}({\bf z}+{\bf p}\tau,\tau) & = & c_{\tau ij}({\bf z},\tau) - p^a c_{ija}({\bf z},\tau) -
\left(p_i p_j + \frac{1}{2} ({\bf p},{\bf p}) g_{ij}\right)\,,\\
c_{\tau\tau i}({\bf z}+{\bf p}\tau,\tau) & = & c_{\tau\tau i}({\bf z},\tau) - 2 p^a c_{\tau ai}({\bf z},\tau) +
p^a p^b c_{abi}({\bf z},\tau)+({\bf p},{\bf p})p_i\,,\\
c_{\tau\tau\tau}({\bf z}+{\bf p}\tau,\tau) & = & c_{\tau\tau\tau}({\bf z},\tau) - 3 p^a c_{\tau\tau a}({\bf z},\tau)+
3 p^a p^b c_{\tau ab}({\bf z},\tau)-p^a p^b p^c c_{abc}({\bf z},\tau)\\&& - \frac{3}{4}({\bf p},{\bf p})^2\,.
\end{eqnarray*}
The $\Delta^{(i)}$ have the following transformation properties:
\begin{eqnarray*}
\Delta^{(3)}_{ijrs}({\bf z}+{\bf p}\tau,\tau) & = &\Delta^{(3)}_{ijrs}({\bf z},\tau)\,,\\
\Delta^{(2)}_{ijk}({\bf z}+{\bf p}\tau,\tau) & = &\Delta^{(2)}_{ijk}({\bf z},\tau)+p^a \Delta^{(3)}_{ijka}({\bf z},\tau)\,,\\
\Delta^{(1)}_{ij}({\bf z}+{\bf p}\tau,\tau) & = &\Delta^{(1)}_{ij}({\bf z},\tau)+
p^a \left\{ \Delta^{(2)}_{ija} ({\bf z},\tau) + \Delta^{(2)}_{jia}({\bf z},\tau)\right\}+
p^a p^b \Delta^{(3)}_{ijab}({\bf z},\tau)\,,\\
\end{eqnarray*}
\end{prop}
We are now in the position to rehearse the main theorem. If we have a full $N$-dimensional
weight lattice, then $\Delta^{(3)}$ is doubly periodic in all ${\bf z}$ variables and if we can show
it has no poles then it must be a function of $\tau$ alone. The modularity properties of
$\Delta^{(3)}$ then imply that it must be zero. Repeating the argument sequentially for $\Delta^{(2)}$ and
then $\Delta^{(1)}$ will give the desired result. To proceed further requires the examination of the
singularity properties of the $\Delta^{(i)}\,.$

\section{Singularity properties}

To study the singularity properties of the $\Delta^{(i)}$ we require a more
detailed analysis of these functions. Using equation (\ref{defF}) and Proposition \ref{wdvvbreakdown}
one obtains:

\begin{eqnarray*}
\Delta^{(1)} & = & \Delta^{(1)}({\bf u},{\bf v})\\
&&\\
 & = & - ({\bf u},{\bf v}) \sum_{\alpha\in\mathfrak{U}} h_\alpha f^{(0,3)}({\bf z}_\alpha,\tau)\\
& &+\sum_{\alpha,\beta\in \mathfrak{U}} h_\alpha h_\beta (\alpha,\beta) (\alpha,{\bf v})
\left[
\begin{array}{c}
+(\beta,{\bf u}) f^{(2,1)}({\bf z}_\beta,\tau) \, f^{(2,1)}({\bf z}_\alpha,\tau) \\
-(\alpha, {\bf u}) f^{(1,2)}({\bf z}_\beta,\tau) \, f^{(3,0)}({\bf z}_\alpha,\tau)
\end{array}
\right]\\
&&\\
&&\\
\Delta^{(2)} & = & \Delta^{(2)}({\bf u},{\bf v},{\bf w}) \\
&&\\
& = & \sum_{\alpha\in\mathfrak{U}} h_\alpha
\left[
({\bf u},{\bf v})(\alpha,{\bf w}) - ({\bf w},{\bf v})(\alpha,{\bf u})
\right] f^{(1,2)}({\bf z}_\alpha,\tau)\\
&&+\sum_{\alpha,\beta\in \mathfrak{U}} h_\alpha h_\beta (\alpha,\beta) (\alpha,{\bf v})
\left[
(\alpha\wedge\beta)({\bf u},{\bf w})
\right]
f^{(2,1)}({\bf z}_\beta,\tau) \, f^{(3,0)}({\bf z}_\alpha,\tau)
\\
&&\\
&&\\
\Delta^{(3)} & = & \Delta^{(3)}({\bf u},{\bf v},{\bf w},{\bf x}) \\
&&\\
& = &
\sum_{\alpha\in\mathfrak{U}}
h_\alpha
\left[
\begin{array}{c}
+(\alpha,{\bf v})(\alpha,{\bf w})({\bf u},{\bf x}) - (\alpha,{\bf x})(\alpha,{\bf w})({\bf u},{\bf v})\\
+(\alpha,{\bf u})(\alpha,{\bf x})({\bf v},{\bf w}) - (\alpha,{\bf u})(\alpha,{\bf v})({\bf w},{\bf x})
\end{array}
\right]
f^{(2,1)} ({\bf z}_\alpha,\tau)\\
&&-\frac{1}{2}
\sum_{\alpha,\beta\in \mathfrak{U}} h_\alpha h_\beta (\alpha,\beta)
[(\alpha\wedge\beta)({\bf u},{\bf w})]
[(\alpha\wedge\beta)({\bf v},{\bf x})]
f^{(3,0)}({\bf z}_\alpha,\tau)f^{(3,0)}({\bf z}_\beta,\tau)\,,
\end{eqnarray*}
where $(\alpha\wedge\beta)({\bf u},{\bf v}) = (\alpha,{\bf u})(\beta,{\bf v}) - (\alpha,{\bf v})(\beta,{\bf u})\,.$

The only derivative of $f$ that gives rise to a pole is the $f^{(3,0)}$ derivative; all other
derivatives are analytic in $z$ - this following from (\ref{fseries1}). Therefore the only parts of the $\Delta^{(i)}$
that could contribute to a singularity are those which contain this derivative.

\begin{prop}\label{poleresult} Let $\Pi_\alpha$ denote a plane through the origin containing the vector $\alpha\,$ and $\alpha^\perp$ a
vector in $\Pi_\alpha$ perpendicular to $\alpha\,.$ Then, at $(\alpha,{\bf z})=0$:

\begin{itemize}
\item{}$\Delta^{(1)}({\bf u},{\bf v})$ has no pole if the scalar equation
\begin{equation}
\label{poleresult1}
\sum_{\beta \in \Pi_\alpha \cap \mathfrak{U}} h_\beta (\alpha,\beta)(\beta,\alpha^\perp)^{2n+1}=0\,, \qquad n=1\,,2\,,\ldots\,,
\end{equation}
holds.

\item{}$\Delta^{(2)}({\bf u},{\bf v},{\bf w})$ has no pole if the bilinear form equation
\begin{equation}
\label{poleresult2}
\sum_{\beta \in \Pi_\alpha \cap \mathfrak{U}} h_\beta (\alpha,\beta) (\alpha \wedge \beta) (\beta,\alpha^\perp)^{2n}=0\,, \qquad n=1\,,2\,,\ldots\,,
\end{equation}
holds.

\item{}$\Delta^{(3)}({\bf u},{\bf v},{\bf w},{\bf x})$ has no pole if the 4-linear form equation
\begin{equation}
\label{poleresult3}
\sum_{\beta \in \Pi_\alpha \cap \mathfrak{U}} h_\beta (\alpha,\beta) (\alpha \wedge \beta)\otimes (\alpha \wedge \beta)(\beta,\alpha^\perp)^{2n+1}=0\,, \qquad n=1\,,2\,,\ldots\,,
\end{equation}
holds.

\end{itemize}
\noindent Here
\[
(\alpha\wedge\beta)({\bf u},{\bf v}) = (\alpha,{\bf u})(\beta,{\bf v}) - (\alpha,{\bf v})(\beta,{\bf u})\,
\]
and
\[(\alpha\wedge\beta)\otimes(\alpha\wedge\beta) ({\bf u},{\bf v},{\bf w},{\bf z})=
\left[(\alpha\wedge\beta)({\bf u},{\bf v})\right]\left[(\alpha\wedge\beta)({\bf w},{\bf x})\right]\,.
\]
\end{prop}

{\sl Proof} The only third derivative of $f$ which contains a pole is the $f^{(3,0)}$-derivative and hence the only part of $\Delta^{(1)}$ that could contain poles is the term
\[
\sum_{\alpha,\beta\in\mathfrak{U}} h_\alpha h_\beta (\alpha,\beta)(\alpha,{\bf v})(\alpha,{\bf u}) f^{(1,2)}({\bf z}_\beta) f^{(3,0)} ({\bf z}_\alpha)\,.
\]
Since $f^{(3,0)}$ only has a simple pole the term involving the poles is, up to a non-zero constant,
\[
\sum_{\alpha\in\mathfrak{U}} \frac{h_\alpha (\alpha,{\bf u})(\alpha,{\bf v})}
{({\bf z},\alpha)}
\sum_{\beta\in\mathfrak{U}} h_\beta (\alpha,\beta) f^{(1,2)}({\bf z}_\beta)\,.
\]
The function $f^{(1,2)}({\bf z}_\beta)$ is odd and hence may be written as $\sum_{n=0}^\infty A_n(\tau) ({\bf z},\beta)^{2n+1}$ (the explicit
expressions for the non-zero functions $A_n$ are not required - they may be derived from Proposition \ref{trans}).
Thus a sufficient condition for the absence of poles is, for arbitrary $\alpha \in \mathfrak{U}\,:$
\[
(\alpha,{\bf z}) {\rm~~divides~~} \sum_{\alpha~{\rm fixed,}~\beta\in\mathfrak{U}} h_\beta (\alpha,\beta) (\beta,{\bf z})^{2n+1}\,, \qquad \qquad n=0\,,1\,,\ldots\,.
\]
Note that this is automatically satisfied if $n=0$ by the first condition in Definition \ref{bigvee}.
This sum may be rewritten as sums over vectors in 2-planes $\Pi_\alpha$ containing $\alpha\,,$ and hence a sufficient
condition for the absence of poles is,
for arbitrary $\alpha \in \mathfrak{U}\,:$
\[
(\alpha,{\bf z}) {\rm~~divides~~} \sum_{\beta \in \Pi_\alpha \cap\mathfrak{U}} h_\beta (\alpha,\beta) (\beta,{\bf z})^{2n+1}\,, \qquad \qquad n=1\,,2\,,\ldots\,.
\]

On  decomposing each $\beta$ in the plane $\Pi_\alpha$ as $\beta = \mu \alpha + \nu \alpha^\perp$ (so $\nu = (\beta,\alpha^\perp)/(\alpha^\perp,\alpha^\perp)$)
one finds that all terms in the binomial expansion of $(\beta,{\bf z})^{2n+1}$ contain a $(\alpha,{\bf z})$-term except the final
$[\nu ({\bf z},\alpha^\perp)]^{2n+1}$-term. Thus a sufficient condition condition for the absence of poles in $\Delta^{(1)}$ is
\[
\sum_{\beta \in \Pi_\alpha \cap \mathfrak{U}} h_\beta (\alpha,\beta)(\beta,\alpha^\perp)^{2n+1}=0\,, \qquad n=1\,,2\,,\ldots\,.
\]
The proof of the $\Delta^{(2)}$ condition is identical: $f^{(2,1)}$ is an even function, and the lowest
term vanishes on using the first condition in Definition \ref{bigvee}.

The function $\Delta^{(3)}$ contains a term
\[
\sum_{\alpha\,,\beta\in\mathfrak{U}} h_\alpha h_\beta (\alpha,\beta)
\left[ (\alpha \wedge \beta)({\bf u},{\bf w}) \right] \,
\left[ (\alpha \wedge \beta)({\bf v},{\bf x}) \right] \, \frac{1}{(\alpha,{\bf z})} \, \frac{1}{(\beta,{\bf z})} \,.
\]
This vanishes by definition of a complex Euclidean $\vee$-system \cite{FV2}\,. The proof of the remaining $\Delta^{(3)}$ condition is identical to the above: $f^{(3,0)}$ is an odd function, and the lowest
term vanishes on using a polarized version of condition (\ref{conditionA})\,.

$~~\Box$

\section{The Main Theorem}

We can now draw the various components together, but first we
define an elliptic $\vee$-system.

\begin{defn} Let $ \mathfrak{U}$ be a complex Euclidean $\vee$-system. An elliptic $\vee$-system is a complex Euclidean $\vee$-system with
the following additional conditions:
\begin{itemize}
\item{} $\sum_{\alpha \in\mathfrak{U}} h_{\alpha} (\alpha,{\bf z})^4 = 3({\bf z},{\bf z})^2\,;$
\item[]
\item{} The three conditions in Proposition \ref{poleresult} hold;
\item[]
\item{} There exists a full $N$-dimensional weight lattice of vectors ${\bf p}$ such that
$({\bf p},\alpha) \in\mathbb{Z}$ for all $\alpha\in \mathfrak{U}\,.$
\end{itemize}
\end{defn}
\noindent Examples of elliptic $\vee$-systems will be constructed in the next section. With this
definition in place one arrives at the main theorem.

\begin{thm} Let $\mathfrak{U}$ be an elliptic $\vee$-system. If $h^\vee_\mathfrak{U}=0$ then the function
\begin{equation}
\label{F}
F(u,{\bf z}, \tau) = \frac{1}{2} u^2 \tau - \frac{1}{2} u ({\bf z},{\bf z}) +  \sum_{\alpha\in\mathfrak{U}} h_{\alpha} f\left(\, {\bf z}_\alpha, \tau\right)
\end{equation}
satisfies the WDVV equations. If $h^\vee_\mathfrak{U}\neq 0$ then the modified prepotential
\begin{equation}
\label{modifiedF}
F\longrightarrow F+ \frac{10 \left( h^\vee_\mathfrak{U}\right)^2 }{3(2\pi i)^3} \mathcal{L}i_3(1,q)
\end{equation}
satisfies the WDVV equations.
\end{thm}
{\sl Proof} From the conditions in the definition of an elliptic $\vee$-system and Proposition \ref{periodprop} it follows that
$\Delta^{(3)}$ is doubly periodic in all ${\bf z}$-variables, and from the conditions in Proposition \ref{poleresult}
it follows that it has no poles. It therefore must be a function of $\tau$ alone. From Proposition \ref{modularprop} it follows
that $\Delta^{(3)}$ is a modular function of degree 2 and from Proposition \ref{modularprop} it follows that it contains only positive powers in its $q$-expansion. Hence it is a modular form of degree 2 and hence must be zero.

This argument can now be repeated for $\Delta^{(2)}$ (a modular function of degree 3 with only positive powers in its $q$-expansion and hence
a modular form of degree 3 and so must be zero).

Finally, the same arguments implies that $\Delta^{(1)}$ is a modular form of degree 4, and hence it must be a multiple of the modular form $E_4(\tau)\,.$ Thus
\[
\Delta^{(1)}({\bf u},{\bf v}) = m({\bf u},{\bf v}) \, E_4(\tau)\,.
\]
To find $m({\bf u},{\bf v})$ one just requires the $O(1)$-terms in the $q$-expansion of $\Delta^{(1)}\,.$ On using equation (\ref{fseries2})
one finds that
\begin{eqnarray*}
m({\bf u},{\bf v})& = & \sum_{\alpha,\beta\in\mathfrak{U}} h_\alpha h_\beta (\alpha,\beta)(\alpha, {\bf v})(\beta,{\bf u})\left(-\frac{1}{12}\right)^2\,,\\
& = & \frac{1}{36} \left( h^\vee_\mathfrak{U}\right)^2 \, ({\bf u},{\bf v}).
\end{eqnarray*}
Hence
\[
\Delta^{(1)} = \frac{1}{36} \left( h^\vee_\mathfrak{U}\right)^2 \, E_4(\tau)\,({\bf u},{\bf v}).
\]
Thus if $h^\vee_\mathfrak{U}=0$ then $\Delta^{(i)}=0$ for $i=1\,,2\,,3$ and hence (\ref{F}) satisfies the WDVV equations.

If $h^\vee_\mathfrak{U}\neq 0$ one has to modify the ansatz for $F\,:$
\[
F\longrightarrow F + \mu \frac{1}{(2 \pi i)^3} \mathcal{L}i_3(1,q)\,.
\]
This change only effects $c_{\tau\tau\tau}$ and hence the above argument on the vanishing of $\Delta^{(3)}$ and $\Delta^{(2)}$ is unchanged.
With this new ansatz $\Delta^{(1)}$ undergoes a slight change:
\[
\Delta^{(1)} \longrightarrow \Delta^{(1)} - \mu \frac{1}{120} E_4(\tau) \, ({\bf u},{\bf v})\,,
\]
on using (\ref{e4int}). Thus if
\[
\mu = \frac{10}{3} \left( h^\vee_\mathfrak{U} \right)^2
\]
then the modified $\Delta^{(1)}$ is zero and hence (\ref{modifiedF}) satisfies the WDVV equations.
$~~\Box$

\subsection{Rational and Trigonometric Limits}\label{triglimits}

From the leading order behaviour, obtained from Proposition \ref{trans}\,,
\[
f=-\frac{1}{(4\pi i)} z^2 \log z\qquad {\rm as~}z \rightarrow 0\,,\\
\]
and
\[
f\simeq\frac{1}{(2\pi i)^3} Li_3\left( e^{2\pi i z} \right) + \frac{1}{12} z^3\qquad {\rm as~}q \rightarrow 0\,
\]
one may obtain rational and trigonometric solutions, of lower dimension, of the WDVV equations.

\begin{prop}
Given an elliptic $\vee$-system $\mathfrak{U}$ the following are solutions of the WDVV equations:

\medskip

\noindent{\underline{Rational limit}}
\[
F^{rational} = \sum_{\alpha\in\mathfrak{U}} h_\alpha (\alpha,{\bf z})^2 \log (\alpha,{\bf z})\,.
\]
The metric in this case is the standard Euclidean inner product on $\mathfrak h\,.$

\medskip

\noindent{\underline{Trigonometric limit I}}

\smallskip

If $h^\vee_\mathfrak{U}=0$ then
\[
F^{trig}=\sum_{\alpha\in\mathfrak{U}} h_\alpha Li_3\left( e^{2\pi i ({\bf z},\alpha)} \right)\,.
\]
The metric in this case is the standard Euclidean inner product on $\mathfrak h\,.$

\medskip

\noindent{\underline{Trigonometric limit II}}

\smallskip

If $h^\vee_\mathfrak{U}\neq0$ then
\[
F^{trig}=\frac{1}{6}u^3 - \frac{1}{2} u ({\bf z},{\bf z}) + \frac{1}{(2\pi i)^3} \left({\frac{3}{h^\vee_\mathfrak{U}}}\right)^{\frac{1}{2}} \sum_{\alpha\in\mathfrak{U}} h_\alpha Li_3\left( e^{2\pi i ({\bf z},\alpha)} \right)\,.
\]
In this case one has a covariantly constant unity vector field $\partial_u$ and hence the metric in this case $g=du^2 - (d{\bf z},d{\bf z})\,.$

\end{prop}

The proof just involves the examination of the associator $\Delta^{(3)}$ under the above mentioned limits.

\section{Examples of elliptic $\vee$-systems}

In this section we construct examples of elliptic $\vee$-systems based on a Weyl group $W$.
Recall that by assumption, if $\alpha\in\mathfrak{U}$ then $-\alpha\in\mathfrak{U}$. We now also assume that the constants
$h_\alpha$ are Weyl invariant, i.e. $h_{w(\alpha)}=h_\alpha$ for $w\in W\,.$ We denote the number of vectors in $\mathfrak{U}$
by $|\mathfrak{U}|\,.$ The calculations for specific groups will be done using the standard notion for roots and weights,
see for example \cite{Humphreys}.

Two classes of examples will be given, the first where $\mathfrak{U}=\mathcal{R}_W$ (where $\mathcal{R}_W$ is the root system of $W$)
and the second where $\mathfrak{U}=\mathcal{R}_W \cup \mathcal{R}_W^{irreg}\,,$ where $\mathcal{R}_W^{irreg}$
contains a set of $W$-invariant vectors that form an irregular orbit under the action of $W\,.$
We first construct $W$-invariant sets of vectors (and constants $h_\alpha$) satisfying the two conditions
\begin{eqnarray}\label{mainconditions1}
\sum_{\alpha \in\mathfrak{U}} h_\alpha (\alpha,{\bf z})^4 & = & 3 ({\bf z},{\bf z})^2 \,, \\
\label{mainconditions2}
\sum_{\alpha \in\mathfrak{U}} h_\alpha (\alpha,{\bf u}) (\alpha,{\bf v}) & = & 2 h^\vee_{\mathfrak{U}} ({\bf u},{\bf v}) \,
\end{eqnarray}
and then check that the conditions in Proposition \ref{poleresult} are satisfied, which will be done
with the help of the following lemma. Recall that these conditions involve summing over vectors in the
plane $\Pi_\alpha \cap \mathfrak{U}\,.$ In the cases to be discussed here these vectors occur in pairs, related by certain
reflections, and the corresponding terms in the sum cancel. Let $\sigma_\alpha \beta$ denote the reflection of the
vector $\beta$ in the line with normal vector $\alpha\,.$ The pairs in the set of vectors $\Pi_\alpha \cap \mathfrak{U}\,$
will occur in two types:

\begin{picture}(100,120)(-35,0)
\put(23,72) {\vector(2,1){40}}
\put(23,72) {\vector(-2,1){40}}
\put(23,72) {\vector(3,0){60}}
\put(-40,90) {$\sigma_\alpha\beta$}
\put(68,90) {$\beta$}
\put(90,70) {$\alpha$}
\put(15,20) {Type A: $\alpha \in \mathcal{R}_W$}
\put(173,72) {\vector(2,1){40}}
\put(173,72) {\vector(2,-1){40}}
\put(173,72) {\vector(3,0){70}}
\put(218,50) {$\sigma_{\beta^\perp}\alpha$}
\put(218,90) {$\alpha$}
\put(250,70) {$\beta=\alpha+\sigma_{\beta^\perp}\alpha$}
\put(205,20) {Type B}
\end{picture}

\noindent Type A pairs are very familiar: they occur in Weyl group (indeed, Coxeter group) root systems (with certain special
angles). Type B pairs will
occur when an extra set of Weyl invariant vectors is appended to the root system  - see Section \ref{extravectors}. Both
these types of configuration appear in $\vee$-systems and deformed root systems \cite{ChalVes, FV,FV2,Veselov}.

\begin{lem}\label{typeAB} Let $\alpha \in \mathfrak{U}$ and suppose that the terms in the sums ${\sum_{\beta\in \Pi_\alpha \cap\mathfrak{U}}}$
occur in pairs of Type A or Type B. Then the conditions in Proposition \ref{poleresult} are satisfied:

\smallskip

\begin{itemize}

\item[(a)] for type A configurations if and only if $h_\beta = h_{\sigma_\alpha \beta}\,;$

\item[]

\item[(b)] for type B configurations if and only if $(\alpha,\beta) h_\beta = (\alpha, \alpha-\beta) h_{\sigma_{\beta^\perp} \alpha}\,.$

\end{itemize}

\end{lem}

{\sl Proof}

Consider the first condition in Proposition \ref{poleresult}, namely equation (\ref{poleresult1}), and consider the
partial sums:

\noindent{\underline{Type A}}
\[
\Xi_A=h_\beta (\alpha,\beta) (\alpha^\perp,\beta)^n + h_{\sigma_\alpha \beta} (\alpha, \sigma_\alpha \beta) (\alpha^\perp,\sigma_\alpha \beta)^n\,;
\]

\noindent{\underline{Type B}}
\[
\Xi_B=h_\beta (\alpha,\beta) (\alpha^\perp,\beta)^n +h_{\sigma_{\beta^\perp} \alpha} (\alpha, \sigma_{\beta^\perp} \alpha) (\alpha^\perp, \sigma_{\beta^\perp} \alpha)^n\,.
\]

\noindent It is easy to show that $\Xi_A=0$ if and only if $h_\beta = h_{\sigma_\alpha \beta}\,.$ Similarly one may show (and here the condition that $\beta=\alpha+\sigma_{\beta^\perp}\alpha$
is used) that $\Xi_B=0$ if and only if
\begin{equation}
(\alpha,\beta) h_\beta = (\alpha, \alpha-\beta) h_{\sigma_{\beta^\perp} \alpha}
\label{typeBcondition}
\end{equation}
The full sum is made up of sums of such paired-terms, and hence is zero. Repeating the argument for the terms that appear in  equations (\ref{poleresult2}) and (\ref{poleresult3}) yields no further conditions.

$~~\Box$

\noindent Note that we have assumed that the $h_\alpha$ are Weyl invariant and hence for type A configurations the conditions in Proposition \ref{poleresult} are automatically satisfied with no extra conditions.

\medskip

To illustrate this we begin with the simplest case, where $W=A_1\,,$ which will reproduce the examples
constructed earlier.

\begin{example} $W=A_1$
\begin{itemize}
\item{${|\mathfrak{U}|=2}\,$}

Let $\mathfrak{U}=\mathcal{R}_{A_1}=\{\pm\alpha\}$ (normalized so that $(\alpha,\alpha)=2$).
Then conditions (\ref{mainconditions1}) and (\ref{mainconditions2}) imply that
\[
h_\alpha=\frac{3}{8}\,,\qquad\qquad h^\vee_{A_1} = \frac{3}{4}
\]
(note their ratio is 2, which is the (dual) Coxeter number of $A_1$). The pole conditions are vacuous in this case. This gives
solution (\ref{A1solutionB}).

\item{$|\mathfrak{U}|=4\,$}

Let $\mathfrak{U}=\{\pm\alpha,\pm{\tilde\alpha}\}$ with $(\alpha,\alpha)=2\,,({\tilde\alpha},{\tilde\alpha})=\nu\,.$
Then conditions (\ref{mainconditions1}) and (\ref{mainconditions2}) imply that
\begin{eqnarray*}
8 h_\alpha + 2 \nu^2 h_{\tilde\alpha} & = & 3 \,, \\
2 h_\alpha + \nu h_{\tilde\alpha} & = & h^\vee_{\mathfrak{U}}\,.
\end{eqnarray*}
Without loss of generality let $h_{\alpha}=\frac{1}{2}\,.$ Then
\[
h_{\tilde\alpha}=-\frac{1}{2\nu^2}\,,\qquad\qquad h^\vee_{\mathfrak{U}} = 1-\frac{1}{2\nu}\,.
\]
Again the pole conditions are vacuous.
The choice $\nu=\frac{1}{2}$ is special $(h^\vee_{\mathfrak{U}} =0)$ and leads to the solution obtained in
Corollary \ref{A1solutionA}.

\end{itemize}

\end{example}

\subsection{The case $\mathfrak{U}=\mathcal{R}_W$}

\[\]
In this case it follows from general theory that (\ref{mainconditions2}) is satisfied for all Weyl groups (if $h_\alpha=1$ for all roots then
$h^\vee_\mathfrak{U}$ is just the dual Coxeter number of $W$).
Since the quartic expression $\sum h_\alpha (\alpha,{\bf z})^4$ is Weyl invariant, by Chevellay's Theorem  (Theorem \ref{chevalley}) it may be
written in terms of fundamental invariant polynomials of degree 2 and degree 4, i.e.
\[
\sum h_\alpha (\alpha,{\bf z})^4 = A \left[ s_2({\bf z}) \right]^2 + B s_4 ({\bf z})
\]
{\em if} such polynomials exist. The quadratic polynomial $s_2$ exists for all groups $W\,;$ one may take $s_2({\bf z}) = ({\bf z},{\bf z})\,.$
Invariant polynomials of degree 4 do not exist for $W=A_2\,,E_{6,7,8}\,,F_4\,,G_2\,.$ Thus for these groups it follows immediately that
(\ref{mainconditions1}) is satisfied. By direct calculation one may show that for the remaining Weyl groups, $W=A_{N\geq 3}\,,B_N\,,D_N$
(where such an invariant polynomial does exist) condition (\ref{mainconditions1}) fails, except for $B_2$ where it holds if a specific
relationship between $h^{(long)}$ and $h^{(short)}$ exists. Thus, in general, for the three infinite families of groups, condition (\ref{mainconditions1}) fails and one has to append an extra set of Weyl-group invariant vectors in order to satisfy this condition: this will be done in the next section.

Since the constants $h_\alpha$ are Weyl invariant the analysis decomposes into cases labeled by the number of independent
Weyl orbits:

\begin{itemize}

\item[$\bullet$]For $W=A_2\,,E_6\,,E_7\,,E_8$ one has a single Weyl orbit, so the constants $h_\alpha$ are all identical. The values of this
constant, and the constant $h^\vee_\mathfrak{U}$ are tabulated below:

\begin{center}
\begin{tabular}{c|cccc}
{\rm Weyl~group~} & $A_2$ & $E_6$ & $E_7$ & $E_8$ \\
\hline
&&&&\\
$h_\alpha$ & $\displaystyle{\frac{1}{3}}$ & $\displaystyle{\frac{1}{6}}$ & $\displaystyle{\frac{1}{8}}$ & $\displaystyle{\frac{1}{12}}$ \\
&&&&\\
$h^\vee_\mathfrak{U}$ & $1$ & $2$ & $\displaystyle{\frac{9}{4}}$ & $\displaystyle{\frac{5}{2}} $
\end{tabular}
\end{center}
(note, $h^\vee_\mathfrak{U}/h_\alpha$=(dual) Coxeter number, as required).

\medskip

\item[$\bullet$]For $W=B_2\,,G_2\,,F_4$ one has two Weyl orbits, labeled by short and long roots. By direct computation one finds that conditions (\ref{mainconditions1})
and (\ref{mainconditions2}) are satisfied with the following data:
\begin{center}
\begin{tabular}{c|ccc}
{\rm Weyl group} & $B_2$ & $G_2$ & $F_4$ \\
\hline
&&&\\
$h^{(long)}$ & $\displaystyle{\frac{1}{4}}$& $\displaystyle{\frac{1-h}{18}}$ & $\displaystyle{\frac{3-h}{6}}$\\
&&&\\
$h^{(short)}$ & $1$& $\displaystyle{\frac{3h-1}{6}}$ & $\displaystyle{\frac{2h-3}{3}}$\\
&&&\\
$h^\vee_\mathfrak{U}$ & $\displaystyle{\frac{3}{2}}$ & $h$ & $h$
\end{tabular}
\end{center}

\end{itemize}

\medskip

\noindent It remains now to check the conditions appearing in
Proposition \ref{poleresult}. It is well known that for a root
system $\mathcal{R}_W$ the configurations $\Pi_\alpha \cap
\mathcal{R}_W$ are two dimensional root systems, namely one of
$\mathcal{R}_{A_1 \times A_1}\,,
\mathcal{R}_{A_2}\,,\mathcal{R}_{B_2}$ or $\mathcal{R}_{G_2}\,,$
and all of these configurations are of type A. Hence by Lemma
\ref{typeAB} these are elliptic $\vee$-systems and hence provide
solutions of the WDVV equations.

\subsection{The case $\mathfrak{U}=\mathcal{R}_W \cup \mathcal{R}_W^{irreg}$}\label{extravectors}

We now turn our attention to the three infinite families, where one has to append an extra set of vectors to the standard
roots in order to satisfy condition (\ref{mainconditions2}). Note the the Weyl groups $A_1\,, A_2$ and $B_2$ appear to be special
in the sense that there are solutions with both $\mathfrak{U}=\mathcal{R}_W$ {\em and} $\mathfrak{U}=\mathcal{R}_W\cup \mathcal{R}_W^{irreg}\,.$
For $A_{N\geq 3}$ and $B_{N\geq 3}$ condition (\ref{mainconditions2}) fails for $\mathfrak{U}=\mathcal{R}_W$.

\subsubsection{The case $W=A_{N\geq 2}$}\label{Ansolution}
\[\]
Let ${\bf z} = \sum_{i=1}^{N+1} z^i {e}_i\,,$ with $\sum_{i=1}^{N+1} z^i=0\,.$ With the later condition the
following identities immediately follow:
\begin{eqnarray*}
\frac{1}{2} \sum_{i \neq j} (z^i - z^j)^2 - \frac{(N+1)}{2} \sum_i \left\{ (z^i)^2 + (-z^i)^2 \right\} & = & 0\,,\\
\frac{1}{2} \sum_{i \neq j} (z^i - z^j)^4 - \frac{(N+1)}{2} \sum_i \left\{ (z^i)^4 + (-z^i)^4 \right\} & = & 3 \left(\sum_i (z^i)^2 \right)^2\,.
\end{eqnarray*}
From these one may obtain the $\alpha$ and $h_\alpha$ satisfying conditions (\ref{mainconditions1}) and (\ref{mainconditions2}) on using
the standard Euclidean inner product.
Let
\begin{eqnarray*}
\alpha^{(ij)} & = & e_i - e_j \,,\\
\beta^{(i)} & = & \displaystyle{\frac{1}{(N+1)} \left( N e_i - \sum_{j \neq i} e_j\right)}\,
\end{eqnarray*}
(so $(\alpha^{(ij)}, {\bf z}) = z^i-z^j$ and $(\beta^{(i)} ,{\bf z})=z^i$). Note both these vectors lie on the hyperplane $\sum_{i=1}^{N+1} z^i=0\,.$ With these it follows that $\mathfrak{U}=\mathcal{R}_{A_N} \cup \mathcal{R}_{A_N}^{irreg}$ where:
\[
\begin{array}{lll}
\mathcal{R}_{A_N}=\{ \alpha^{(ij)}\,,i \neq j\}\,, & & h_\alpha=1/2 {\rm ~if~}\alpha \in \mathcal{R}_{A_N}\,;\\
&&\\
\mathcal{R}_{A_N}^{irreg}=\{ \pm \beta^{(i)},\,,i=1\,,\ldots\,N+1 \}\,, & & h_\alpha=-(N+1)/2 {\rm ~if~}\alpha \in \mathcal{R}_{A_N}^{irreg}\,.
\end{array}
\]
Note that $\mathcal{R}_{A_N}$ is just the root system for $A_N\,.$ The geometry of this configuration will now be
discussed.

Let $\sigma_\alpha \beta$ denote the reflection of $\beta$ in the plane perpendicular to $\alpha\,.$ Then
\begin{eqnarray*}
\sigma_{\alpha^{(ij)}} \beta^{(i)} & = & \beta^{(i)} - \frac{2(\alpha^{(ij)},{\beta^{(i)}})}{(\alpha^{(ij)},\alpha^{(ij)})} \, \alpha^{(ij)}\,,\\
& = & \beta^{(i)} - \alpha^{(ij)}\,,\\
& = & \beta^{(j)}\,,\\
\sigma_{\alpha^{(ij)}} \beta^{(k)} & = & \beta^{(k)} \,, \quad{i,j,k~{\rm  distinct}}\,.
\end{eqnarray*}
Thus the set $\mathcal{R}_{A_N}^{irreg}$ is invariant under the action of $W$ (which is generated by reflections defined by the vectors in
$\mathcal{R}_{A_N}$).  Thus for $N\geq 3$ the Weyl orbit of an element of $\mathcal{R}_{A_N}^{irreg}$ is smaller  (since $|\mathcal{R}_{A_N}^{irreg}|=2(N+1)$) than the size of the orbit of a generic vector (which would be $|{A_N}|=(N+1)!).$ Thus $\mathcal{R}_{A_N\geq 2}^{irreg}$ consists of the union of two irregular orbits
\[
\mathcal{R}_{A_N\geq 2}^{irreg} \cong \{+\beta^{(i)} | i=1\,, \ldots\,,N+1\} \cup \{-\beta^{(i)} | i=1\,, \ldots\,,N+1\}\,.
\]

There are certain degeneracies if $N=1$ or $2\,:$  if $N=1$ then $\beta^{(1)}=-\beta^{(2)}$ and hence the set $\{\pm\beta^{(i)}\}$ double counts the vectors (this degeneracy was removed in the earlier discussion of
the $A_1$ solution); if $N=2$ then $|\mathcal{R}_{A_N}^{irreg}|=|\mathcal{R}_{A_N}|=(N+1)!\,.$ In fact this case coincides with the $G_2$ example above, i.e.
$\mathcal{R}_{G_2}  \cong \mathcal{R}_{A_2} \cup \mathcal{R}_{A_2}^{irreg}\,$ for a specific value of the constant $h\,,$ namely $h=0\,.$

Note that since $(\beta^{(i)},\alpha^{(jk)})=0$ and $(\beta^{(i)},\alpha^{(ij)})=1$ the set $\mathcal{R}_{A_N}^{irreg}$ consists of vectors
from the weight lattice of $A_N\,.$ In terms of fundamental weights
\[
\Delta_{(i)}=\sum_{r=1}^i e_r - \frac{i}{(N+1)} \sum_{r=1}^{N+1} e_r
\]
one has
\[
\mathcal{R}_{A_N}^{irreg} \cong \left\{ \pm w(\Delta_{(N)})\,:\,w \in W \right\}\,
\]
(note that $\pm\Delta_{(1)}$ also lie in these two orbits). The orbits of other fundamental weights form other
irregular orbits.

Furthermore, if $N\geq 3$ one obtains the configurations
\begin{eqnarray*}
\mathfrak{U}\cap {\rm Span}\{ \alpha^{(ij)},\alpha^{(rs)} \} & = & \mathcal{R}_{A_1 \times A_1} \,, \quad \{i,j\} \cap \{r,s\} \neq \emptyset\,,\\
\mathfrak{U}\cap {\rm Span}\{ \alpha^{(ij)},\alpha^{(ik)} \} & = & \mathcal{R}_{A_2} \,, \quad {i,j,k} {\rm ~distinct}\,
\end{eqnarray*}
together with the new configuration
\[
\mathfrak{U}\cap {\rm Span}\{ \beta^{(i)}\,,\beta^{(j)} \}  =  \{ \pm \beta^{(i)}\,, \pm \beta^{(j)}\,, \pm \alpha^{(ij)} \}\,.
\]
The geometry of this new configuration is shown in Figure 1.
\begin{figure}
\begin{picture}(100,120)(0,0)
\put(51,78) {$\theta$}
\put(53,72) {\vector(2,-1){40}}
\put(53,72) {\vector(-2,-1){40}}
\put(53,72) {\vector(2,1){40}}
\put(53,72) {\vector(-2,1){40}}
\put(53,72) {\vector(3,0){80}}
\put(53,72) {\vector(-3,0){80}}
\put(0,40) {$-\beta^{(i)}$}
\put(0,100) {$+\beta^{(j)}$}
\put(88,40) {$-\beta^{(j)}$}
\put(88,100) {$+\beta^{(i)}$}
\put(135,70) {$\alpha^{(ij)}=\beta^{(i)}-\beta^{(j)}$}
\put(-110,70) {$\alpha^{(ji)}=\beta^{(j)}-\beta^{(i)}$}
\put(25,20) {$\cos\theta = -\frac{1}{N}$}
\end{picture}
\caption{The configuration $\mathfrak{U}_{A_N}\cap {\rm Span}\{ \beta^{(i)}\,,\beta^{(j)} \} $}
\end{figure}
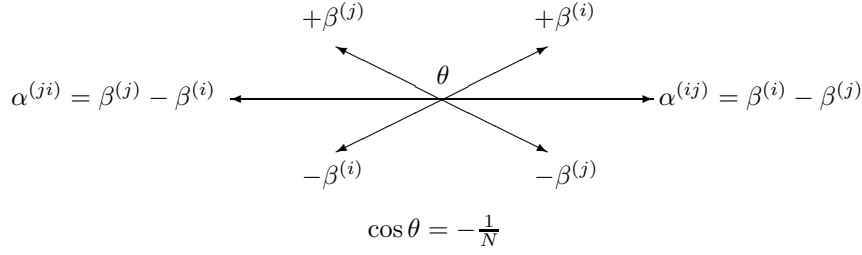
This is precisely a type B configuration, and the condition
(\ref{typeBcondition}) is satisfied, since $\alpha=\beta^{(i)}\,,
h_\alpha=-(N+1)/2$ and $\beta=\alpha^{(ij)}\,, h_\beta=1/2.$ Hence
by Lemma \ref{typeAB} we have an elliptic $\vee$-system and
hence a solution to the WDVV equations.

\subsubsection{The case $W=B_{N}$}\label{Bnsolution}
\[\]
The dual prepotential for the Jacobi group orbit space $\Omega/J(B_N)$ may be calculated in the same was as the
$\Omega/J(A_N)$ dual prepotential was derived in Theorem \ref{RileyStrachan} (see also Example \ref{anbnexample}), and
from this the set $\mathfrak{U}$ and the constants $h_\alpha$ may be extracted.

Given this origin of the set one might expect that it should be related to the root system $\mathcal{R}_{B_N}\,.$ It turns out that
one may describe this set in two ways: either in terms of the root system $\mathcal{R}_{BC_N}$ or in terms of the root system
$\mathcal{R}_{C_N}$ (which is, of course, dual to the root system $\mathcal{R}_{B_N}$) together with an irregular orbit
$\mathcal{R}^{irreg}_{C_N}\,.$

Consider the following identities\footnote{Note the condition $\sum z_i=0$ used in the last section is not used in this section.}:
\begin{eqnarray*}
\sum_{i\neq j} (z^i-z^j)^2 + (z^i+z^j)^2 +\sum_{i=1}^N (2z^i)^2  - 4N \sum_{i=1}^N (2z^i)^2 & = & 0\,,\\
\sum_{i\neq j} (z^i-z^j)^4 + (z^i+z^j)^4 +\sum_{i=1}^N (2z^i)^4  - 4N\sum_{i=1}^N  (2z^i)^4 & = & 12 \left(\sum_i (z^i)^2 \right)^2\,.
\end{eqnarray*}
On defining the inner product to be twice the standard Euclidean product (that is, $({\bf z},{\bf z})=2\sum_i (z^i)^2$) one  may obtain the $\alpha$ and $h_\alpha$ satisfying conditions (\ref{mainconditions1}) and (\ref{mainconditions2}).

In terms of the root system $\mathcal{R}_{BC_N}\,,$ one has $\mathfrak{U}_{B_N} = \mathcal{R}_{BC_N}$ where
\[
\mathcal{R}_{BC_N} = \left\{ \frac{1}{2} (\pm e_i\pm e_j)\,,i \neq j\right\} \cup \left\{ \pm e_i\right\}\cup \left\{ \pm \frac{1}{2} e_i\right\}
\]
and
\[
h_{\alpha} = \left\{
\begin{array}{ccl}
\frac{1}{2} & &{\rm if}~\alpha~{\rm is~a~long~root\,,}\\
1 & &{\rm if}~\alpha~{\rm is~a~middle~root\,,}\\
-2N & &{\rm if}~\alpha~{\rm is~a~short~root\,.}
\end{array}\right.
\]
Alternatively (and this provides a description that is closer to the $A_N$ configuration above)
\begin{eqnarray*}
\mathfrak{U} & = & \mathcal{R}_{C_N} \cup \mathcal{R}_{C_N}^{irreg}\,,\\
& = & \mathcal{R}^{\vee}_{B_N} \cup \mathcal{R}_{B_N}^{\vee\,\,irreg}
\end{eqnarray*}

where:
\[
\begin{array}{lll}
\mathcal{R}_{C_N}=\{ \frac{1}{2} (\pm e_i\pm e_j)\,,i \neq j\} \cup \{ \pm e_i\} \,, & & h_\alpha=
\left\{
\begin{array}{ll}
1&{\rm~if~}\alpha~{\rm short}\\
1/2& {\rm ~if~}\alpha~{\rm long}
\end{array}
\right\}~{\rm if~}
\alpha \in \mathcal{R}_{C_N}\,;\\
&&\\
\mathcal{R}_{C_N}^{irreg}=\{ \pm \frac{1}{2} e_i \}\,, & & h_\alpha=-2N {\rm ~if~}\alpha \in \mathcal{R}_{C_N}^{irreg}\,.
\end{array}
\]
As in the $A_N$ case, $\mathcal{R}_{C_N}^{irreg}$ is an irregular orbit (a single orbit in this case):
\[
\mathcal{R}_{C_N}^{irreg}= \{ w(\Delta_{(N)}) | w \in W\}
\]
for a certain fundamental weight $\Delta_{(N)}\,.$

In either case, the only new two dimensional configuration on vectors is
$\mathfrak{U}_{C_N}\cap {\rm Span}\{ e_i\,,e_j \}$. This is shown
in Figure 2, where the vectors of $\mathcal{R}_{C_N}^{irreg}$ have
been displaced slightly for visual reasons (this is actually the $BC_2$ system). The proof that this is
an elliptic $\vee$ system follows the $A_N$ case and will be
omitted. It also follows from the Hurwitz space description that will be given in
Section \ref{HurwitzSection}\,.

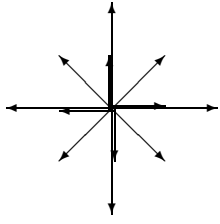
\begin{figure}
\begin{picture}(100,120)(0,0)
\put(53,72) {\vector(2,0){40}}
\put(53,72) {\vector(-2,0){40}}
\put(53,72) {\vector(0,2){40}}
\put(53,72) {\vector(0,-2){40}}
\put(53,72) {\vector(1,1){20}}
\put(53,72) {\vector(-1,-1){20}}
\put(53,72) {\vector(-1,1){20}}
\put(53,72) {\vector(1,-1){20}}
\put(52,73) {\vector(2,0){21}}
\put(54,71) {\vector(-2,0){21}}
\put(52,71) {\vector(0,2){21}}
\put(54,73) {\vector(0,-2){21}}
\end{picture}
\caption{The configuration $\mathfrak{U}_{B_N}\cap {\rm Span}\{ e_i\,,e_j \} $}
\end{figure}

\subsubsection{The case $W=D_N$}
\[\]
The $D_N$ configurations are combinatorially quite complicated, as, even at $N=5$ several Weyl orbits of fundamental weights
have to be appended to the basic root system $\mathcal{R}_{D_N}\,$ in order to satisfy (\ref{mainconditions1}) and (\ref{mainconditions2}).
Some of the resulting configurations $\Pi_\alpha \cap \mathfrak{U}$ are not of Type A and Type B. This does not
mean that the conditions in Proposition \ref{poleresult} must be false - there may be other reasons why the various terms could vanish.

In the $N=4$ case $\mathfrak{U}=\mathcal{R}_{D_4} \cup \mathcal{R}_{D_4}^{irreg}$ which coincides with the $F_4$ example
considered above, with the long roots of $F_4$ being the roots of $D_4$ and the short roots being interpreted as the
irregular orbits of the fundamental weights of $D_4\,.$ Clearly more work
is required to construct an example of a $D_N$ elliptic $\vee$-system.

\bigskip

\noindent The results in this section have been obtained on a case-by-case basis. It would be nice if there was a more
abstract derivation of the results.

\section{Frobenius-Stickelberger Identities}\label{FSconditions}

Hidden within the vanishing of the $\Delta^{(i)}$ are a number of interesting functional
identities satisfied by the various third derivatives of the elliptic trilogarithm, the simplest of these
reducing to $19^{\rm th}$ century $\vartheta$-function identities. We build up to these by first
considering the rational and trigonometric versions.
Given non-zero $a\,,b\,,c \in\mathbb{C}$ such that $a+b+c=0$ then
\[
\frac{1}{a} \,.\,\frac{1}{b} +\frac{1}{b} \,.\,\frac{1}{c} +\frac{1}{c} \,.\,\frac{1}{a} =0
\]
and
\[
\cot(a) \cot(b) +\cot(b) \cot(c) +\cot(c) \cot(a) =1\,.
\]
Such identities are used in the direct verification that the rational (\ref{dualrational})
and trigonometric (\ref{dualtrig}) prepotentials satisfy the WDVV equations. The elliptic version (where the dependence on
$\tau$ has been suppressed for notational convenience) is
\[
\left\{
\begin{array}{c}
\phantom{+}f^{(3,0)}(a) f^{(3,0)}(b)\\
\\
+f^{(3,0)}(b) f^{(3,0)}(c)\\
\\
+f^{(3,0)}(c) f^{(3,0)}(a)
\end{array}
\right\}
-\left\{ f^{(2,1)}(a) + f^{(2,1)}(b) + f^{(2,1)}(c)\right\}=0
\]
Using (\ref{fthetaeta}) this may be written in terms of $\vartheta$-functions\footnote{
This formula was found by the author during the researches that led to \cite{RS} and it has also
appeared recently, with proof, in the work of Calaque, Enriques and Etingof \cite{Etingof}. However it is a classical
formula; in terms of Weierstrass functions it is just the well known Frobenius-Stickelberger
equation \cite{W}
\[
(\zeta(a)+\zeta(b)+\zeta(c))^2 = \wp(a)+\wp(b)+\wp(c)\,,\qquad (a+b+c=0)
\]
re-written in terms of $\vartheta$-functions,
an observation due to Prof. H.W.Braden.}:
\[
\frac{\vartheta_1'(a)}{\vartheta_1(a)}
\frac{\vartheta_1'(b)}{\vartheta_1(b)}+
\frac{\vartheta_1'(b)}{\vartheta_1(b)}
\frac{\vartheta_1'(c)}{\vartheta_1(c)}+
\frac{\vartheta_1'(c)}{\vartheta_1(c)}
\frac{\vartheta_1'(a)}{\vartheta_1(a)}+ \frac{1}{2} \left[
\frac{\vartheta_1''(a)}{\vartheta_1(a)}+
\frac{\vartheta_1''(b)}{\vartheta_1(b)}+
\frac{\vartheta_1''(c)}{\vartheta_1(c)}\right]=
\frac{1}{2} \frac{\vartheta_1'''(0)}{\vartheta_1'(0)}
\]
where $a+b+c=0\,.$
With the
identification $a=(\alpha,{\bf z})\,, b=(\beta,{\bf z})\,,c=-(\alpha+\beta,{\bf z})$ these
identities may be seen as identities connected to the $A_2$ Coxeter group, with $\alpha$
and $\beta$ being the positive roots. This immediately motivates the following:

\begin{lem} Let $\mathcal{R}$ be the root system for the 2-dimensional Coxeter groups
$A_2\,,B_2$ or $G_2\,,$ with the standard normalization
for $\alpha\,,\beta$ positive simple roots:
\begin{eqnarray*}
A_2\,: &&(\alpha,\alpha)=(\beta,\beta)=2\,,(\alpha,\beta)=-1\,,\\
B_2\,: &&(\alpha,\alpha)=2\,,(\beta,\beta)=1\,,(\alpha,\beta)=-1\,,\\
G_2\,: &&(\alpha,\alpha)=6\,,(\beta,\beta)=2\,,(\alpha,\beta)=-3\,.\\
\end{eqnarray*}
Then
\[
\sum_{\alpha\neq\beta\in\mathcal{R}^{+}} (\alpha,\beta) f^{(3,0)}({\bf z}_\alpha,\tau)\,.\,f^{(3,0)}( {\bf z}_\beta,\tau)
+ \sum_{\alpha\in\mathcal{R}^{+}} k_{\alpha} f^{(2,1)}({\bf z}_\alpha,\tau) = 0\,,
\]
where:
\begin{itemize}
\item{$A_2\,:$} $k_\alpha =1$ for all roots;
\item[]
\item{$B_2\,:$} $k_{short}=2\,,k_{long}=1$;
\item[]
\item{$G_2\,:$} $k_{short}=10\,,k_{long}=6$.
\end{itemize}
\end{lem}
The proof is entirely standard and is omitted. Many other functional identities
may be derived using the same ideas. Rather than give a full list we present two of the $A_2$ identities:

\[
\left\{
\begin{array}{c}
\displaystyle{f^{(3,0)}(x + y) \left[f^{(2,1)}(x) - f^{(2,1)}(y)\right]}\\
\\
+\displaystyle{f^{(3,0)}(y)\left[ f^{(2,1)}(x + y) - f^{(2,1)}(x)\right]}
\end{array}
\right\}
+f^{(1,2)}(x) - \frac{1}{2} f^{(1,2)}(y) + \frac{1}{2} f^{(1,2)}(x + y)=0
\]

\noindent and

\begin{eqnarray*}
\left\{
\begin{array}{c}
\phantom{+}\displaystyle{f^{(3,0)}(x)\left[f^{(1,2)}(x + y) - f^{(1,2)}(y)\right]}\\
\\
+\displaystyle{f^{(3,0)}(y)\left[f^{(1,2)}(x + y) - f^{(1,2)}(x)\right]}\\
\\
-\displaystyle{\frac{2}{3}}
\displaystyle{f^{(3,0)}(x + y)\left[f^{(1,2)}(x) + f^{(1,2)}(y)\right]}
\end{array}
\right\}
&+&
\left\{
\begin{array}{c}
\phantom{+}\displaystyle{\frac{2}{3} f^{(2,1)}(x + y)f^{(2,1)}(x)}\\
\\
+\displaystyle{\frac{2}{3} f^{(2,1)}(x + y)f^{(2,1)}(y)}\\
\\
-\displaystyle{\frac{8}{3} f^{(2,1)}(x) f^{(2,1)}(y)}
\end{array}
\right\}\\
&&\\
&&\\
+\frac{10}{9} f^{(0,3)}(x + y) &=& -\frac{1}{108} E_4(\tau)\,.
\end{eqnarray*}
Clearly there is much scope to investigate such neo-classical functional identities. More identities of these type
may be found in \cite{iabs2}\,.

\section{Jacobi Group Orbit Spaces}\label{hurwitzdetails}

Mention has been made a number of times to Jacobi groups and their orbit spaces, but so far
these have not been defined. In this section this is rectified and in addition the construction of the
Frobenius manifold structure on such orbit spaces will be outlined. In particular, using an
alternative description of such spaces as specific Hurwitz spaces we construct the dual
prepotentials for the Weyl groups $A_N$ and $B_N\,,$ thus proving that the examples of
elliptic $\vee$-systems constructed earlier correspond to Jacobi group orbit spaces. This
then motivates a conjecture for arbitrary Weyl group.

\subsection{Jacobi groups and Jacobi forms}

The material in this section will closely follow \cite{B}, which in turn relies heavily on the
fundamental papers of Wirthm\"uller \cite{wirth} and Eichler and Zagier \cite{EZ}. We begin by the definition of a Jacobi form.
These play the same role in the construction of the orbit space as the symmetric polynomials do
in the original Saito construction - they provide coordinates on the orbit space.

\begin{defn}\label{jacformdef}
Let $W$ be a finite Weyl group with root lattice $Q$ and let $\mathfrak{g}$ be the corresponding Lie algebra with Cartan subalgebra $\mathfrak{h}\,.$
A Jacobi form of weight $k\in \mathbb{Z}$ and index $m \in \mathbb{Z}$ is a holomorphic function $\phi:\mathfrak{h}\oplus \mathbb{H} \rightarrow \mathbb{C}$
with the following properties:
\begin{eqnarray*}
\phi({\bf z}+{\bf q},\tau) & = & \phi({\bf z},\tau)\,,\\
\phi({\bf z}+{\bf q}\tau,\tau) & = & e^{-2 \pi i m ({\bf q},{\bf z}) - \pi i m ({\bf q},{\bf q}) \tau}\,.\, \phi({\bf z},\tau)\,,\quad {for~all~} {\bf q}\in Q\,,\\
\phi\left( \frac{{\bf z}}{c \tau + d},\frac{ a \tau+b}{c \tau + d}\right) &= &(c \tau + d)^k \,.\, e^{c \pi i m ({\bf z},{\bf z})/(c \tau+d)}\,.\, \phi({\bf z},\tau)\,,\\
\phi(w.{\bf z},\tau) & = & \phi({\bf z},\tau)\,,\quad {for~all~} w\in W\\
\end{eqnarray*}
and $\phi({\bf z},\tau){~is~a~locally~bounded~function~as~}\Im m(\tau)\rightarrow +\infty\,.$
\end{defn}

Such forms are the elliptic analogues of the $W$-invariant polynomials and they too satisfy a Chevalley-type theorem.

Following Bertola \cite{B} , one can defined a new function $\phi(u,{\bf z},\tau) = e^{mu} \phi({\bf z},\tau)$ defined on the Tits cone
$\Omega\cong \mathbb{C} \oplus \mathfrak{h}\oplus \mathbb{H}\,.$ This is also
referred to as a Jacobi form and the space of Jacobi forms will be denoted $J^W_{k,m}\,.$ The Jacobi group $J(\mathfrak{g})$ itself generates the above transformations.
The full details are not required here: $J(\mathfrak{g})$ is the semi-direct product $\mathcal{W}\rtimes SL(2,\mathbb{Z})$ where
$\mathcal{W}=W \rtimes H_\mathcal{R}$ where $W$ is a Weyl group and $H_\mathcal{R}$ the Heisenberg group obtained from the root space $\mathcal{R}$ of $W\,.$ The
precise definitions of the various actions may be found in \cite{B,wirth}.

It is well known that the ring of modular forms is a free graded algebra over $\mathbb{C}$
generated by the Eisenstein series $E_4$ and $E_6\,,$ i.e. $M_\bullet=\bigoplus_k {\bf M}_k\,,$ where the subspace of modular forms of weight $k$ is
\[
{\bf M}_k = \mathbb{C}[E_4^a E_6^b, \forall a\,,b\,\in \mathbb{N}\, such~that~ 4a+6b=k]\,.
\]
The ring of Jacobi forms is particularly nice; it satisfies an analogue of Chevalley's Theorem (Theorem \ref{chevalley})\,:
\begin{thm}\cite{wirth}
Given the Jacobi group associated to any finite dimensional simple Lie algebra $\mathfrak{g}$ of rank $N$ (except for possibly $E_8$):
\begin{itemize}
\item{} the bi-graded algebra of Jacobi forms $J^W_{\bullet,\bullet}=\bigoplus_{k,m} J_{km}^W$ is freely generated by $N+1$ fundamental
Jacobi forms $\{\phi_0\,,\ldots\,,\phi_N\}$ over the graded ring of modular forms ${\bf M}_\bullet\,,$
\[J^W_{\bullet,\bullet}={\bf M}_\bullet[\phi_0\,,\ldots\,,\phi_N]\,;\]
\item{} each
\[
\phi_j \in J^W_{-k(j),m(j)}
\]
where $-k(j)\leq 0\,,m(j)>0$ are defined as follows:
\begin{itemize}
\item{} $k(0)=1$ and $k(j)\,,j>0$ are the degrees of the generators of the invariant polynomials in
(\ref{chevalley})\,;
\item{} $m(0)=1$ and $m(j)\,,j>0$ are the coefficients in the expansion of the highest coroot ${\tilde \alpha}^{\vee}\,,$
\[
{\tilde \alpha}^{\vee} = \sum_{j=1}^N m(j) \alpha^\vee_j
\]
where ${\tilde\alpha}$ is the highest root and $ \alpha^\vee_j$ a basis for $\mathcal{R}^\vee\,.$
\end{itemize}
\end{itemize}
\end{thm}
Note that $J_{\bullet,0}^{W}\cong{\bf M}_\bullet\,.$ It will also be useful to define $\phi_{-1}=\tau$, even though it is
not a Jacobi form. These Jacobi forms become the coordinates on the orbit space $\Omega/J(\mathfrak{g})\,.$
Before turning to the explicit construction of such forms we prove the following simple
result on the Jacobian of the transformation between the two coordinate systems $\{u,{\bf z},\tau\}$ and
$\{\phi_{-1}\,,\phi_0\,,\ldots\,,\phi_N\}\,.$

\begin{prop}
Let
\[
Jac(u,{\bf z},\tau) = \frac{\partial \{\phi_{-1}\,,\phi_0\,,\ldots\,, \phi_N\}}{\partial\{ u,{\bf z},\tau\}}\,.
\]
Then $Jac$ has the following transformation properties:
\begin{eqnarray*}
\frac{1}{2\pi i} \frac{\partial~}{\partial u} Jac(u,{\bf z},\tau) & = & h^\vee\, Jac(u,{\bf z},\tau)\,,\\
Jac(u,{\bf z}+{\bf q},\tau) & = & Jac(u,{\bf z},\tau)\,,\\
Jac(u,{\bf z}+{\bf q}\tau,\tau) & = & e^{-2\pi i h^\vee ({\bf q},{\bf z}) - \pi i h^\vee ({\bf q},{\bf q})} \,.\,Jac(u,{\bf z},\tau)\,,\\
Jac(u,{\bf z},\tau+1) & = & Jac(u,{\bf z},\tau)\,,\\
Jac\left(u,\frac{{\bf z}}{\tau},-\frac{1}{\tau}\right) & = & \tau^{-|\mathcal{R}^{+}_W|} e^{\pi i h^\vee ({\bf z},{\bf z})/\tau}\,.\,Jac(u,{\bf z},\tau)\,,\\
Jac(u,w.{\bf z},\tau) & = & \det(w)\,.\,Jac(u,{\bf z},\tau)\,,
\end{eqnarray*}
where $h^\vee$ is the dual Coxeter number and $|\mathcal{R}^{+}_W|$ the number of positive roots. Moreover, up to
an overall constant,
\begin{equation}
\label{jac}
Jac(u,{\bf z},\tau) = e^{2 \pi i h^\vee u} \prod_{\alpha \in \mathcal{R}^{+}_W} \frac{\vartheta_1( {\bf z}_\alpha,\tau)}{\vartheta_1^\prime(0,\tau)}\,.
\end{equation}
\end{prop}
\noindent These transformation properties may be elevated to a definition of an anti-invariant Jacobi form. This result is the
elliptic version of the well known result $Jac({\bf z})=\prod_{\alpha \in \mathcal{R}^{+}_W}  {\bf z}_\alpha$ for
Coxeter groups.

{\sl Proof} By definition, the Jacobian is a determinant, so by using properties of the determinant, together with the
transformation properties of the individual Jacobi forms given in Proposition \ref{jacformdef} the result follows. Various Lie-theory results are used, such as
\[
h^\vee = \sum_{i=0}^N m(i)\,,\quad |\mathcal{R}^{+}_W|= \sum_{i=1}^N (k(i)-1)\,,
\]
proofs of which may be found in Kac \cite{Kac}.

To prove (\ref{jac}) one first proves that the right-hand-side has the same transformation properties as $Jac\,.$ Therefore
their ratio transformations like a $J_{0,0}^{W}$-Jacobi form (the analytic properties following from those of the $\vartheta_1$-function,
such as its entire property). But $J_{0,0}^{W} \cong M_0$ and there are no non-trivial degree $0$ modular forms and hence the ratio must be
a constant.
$~~\Box$

Further properties of the forms may be found in \cite{B,wirth}. For the $A_N$ and $B_N$ cases there is a very compact way to study the
forms by combining them into a generating function.
The invariant polynomials for the $A_N$-Coxeter group may be obtained via a generating function (a result due to Vi\`ete)
\[
\left.\prod_{i=0}^N (v-z_i)\right|_{\sum z^i=0} = v^{N+1} +\sum_{r=0}^{N-1} (-1)^{N+1-r} s_{r+1}({\bf z}) v^r\,.
\]
Similarly, the $A_N$ Jacobi forms may be obtained \cite{B} from a similar expansion of
\begin{equation}
\label{Anpotential}
\lambda^{A_N}(v) = e^{2 \pi i u} \displaystyle{\left.\frac{{\prod_{i=0}^N \vartheta_1(v-z_i,\tau)}}{\vartheta_1(v,\tau)^{N+1}}\right|_{\sum z^i=0}}
\end{equation}
as a sum of Weierstrass $\wp$ functions and their derivatives, their coefficients being the $A_N$-Jacobi forms. Using the
embedding $B_N \subset A_{2N-1}$ one may obtain a generating function for the $B_N$-Jacobi forms:
\begin{equation}
\label{Bnpotential}
\lambda^{B_N}(v) = e^{2 \pi i u} \displaystyle{\frac{\prod_{i=1}^N \vartheta_1(v-z_i,\tau) \vartheta_1(v+z_i,\tau)}{\vartheta_1(v,\tau)^{2N}}}
\end{equation}
These generating functions are not just formal objects, they are holomorphic maps from the complex torus to the Riemann sphere.
This means one can use a Hurwitz space construction to calculate the dual prepotential.

\subsection{Hurwitz spaces}\label{HurwitzSection}
Let $H_{g, N}(k_1, \dots, k_l)$ be the Hurwitz space\footnote{
Dubrovin \cite{dubrovin1} uses a slightly different notation. In his notation the
Hurwitz space is $H_{g;k_1-1\,,\ldots\,,k_{l}-1}\,.$} of
equivalence classes $[\lambda:{\mathcal{L}}\rightarrow {\mathbb P}^1]$
of $N$-fold branched coverings $\lambda:{\mathcal{L}}\rightarrow
{\mathbb P}^1$, where $\mathcal{L}$ is a compact Riemann surface
of genus $g$ and the holomorphic map $\lambda$ of degree $N$ is subject
to the following conditions:
\begin{itemize}
\item it has $M$ simple ramification points $P_1, \dots,
P_M\in{\mathcal{L}}$ with distinct {\it finite} images $\l_1,
\dots, \l_M\in {\mathbb C}\subset {\mathbb P}^1$; \item the
preimage $\lambda^{-1}(\infty)$ consists of $l$ points:
$\lambda^{-1}(\infty)=\{\infty_1, \dots,\infty_l\}$, and the
ramification index of the map $\lambda$ at the point $\infty_j$ is $k_j$
($1\leq k_j\leq N$).
\end{itemize}

\noindent (We define the ramification index at a point as the
number of sheets of the covering which are glued together at this
point. A point $\infty_j$ is a ramification point if and only if
$k_j>1$. A ramification point is simple if the corresponding
ramification index equals $2$.) The Riemann-Hurwitz formula
implies that the dimension of this space is $M=2g+l+N-2$. One has
also the equality $k_1+\dots +k_l=N$. Two branched coverings
$\lambda_1:{\mathcal{L}}_1\rightarrow {\mathbb P}^1$ and
$\lambda_2:{\mathcal{L}}_2\rightarrow {\mathbb P}^1$ are said to be
equivalent if there exists a biholomorphic map
$f:{\mathcal{L}}_1\to {\mathcal{L}}_2$ such that $\lambda_2f=\lambda_1$.

We also introduce the covering $\hat{H}_{g, N}(k_1, \dots, k_l)$ of the space $H_{g, N}(k_1,
\dots, k_l)$ consisting of pairs
$$<[\lambda:{\mathcal{L}}\to{\mathbb P}^1]\in H_{g, N}(k_1, \dots, k_l), \{a_\alpha, b_\alpha\}_{\alpha=1}^g>,$$
where $\{a_\alpha, b_\alpha\}_{\alpha=1}^g$ is a canonical basis
of cycles on the Riemann surface $\mathcal{L}$. The spaces
$\hat{H}_{g, N}(k_1, \dots, k_l)$ and $H_{g, N}(k_1, \dots, k_l)$
are connected complex manifolds and the local coordinates on these
manifolds are given by the finite critical values of the map $\lambda\,.$
For $g=0$ the spaces $ \hat{H}_{g,
N}(k_1, \dots, k_l)$ and $H_{g, N}(k_1, \dots, k_l)$ coincide.

The various metric and multiplication tensors are given
in terms of this holomorphic map $\lambda:{\mathcal{L}}\rightarrow {\mathbb P}^1$ (also
known as the superpotential) by the following:

\begin{thm}\label{superpotential}
The intersection form and dual multiplication
on the Hurwitz space $H_{g, N}(k_1, \dots, k_l)$ are given by the
following residue formulae:
\begin{eqnarray*}
g(\partial',\partial{''}) & = &  \sum \res_{d\lambda=0}
\frac{\partial'(\log\lambda(v) dv) \partial{''}(\log\lambda(v)
dv)}{d\log\lambda(v)} \,, \\
{c}^\star\,(\partial',\partial{''},\partial{'''}) & = & \frac{1}{2 \pi i} \sum
\res_{d\lambda=0} \frac{\partial'(\log\lambda(v) dv)
\partial{''}(\log\lambda(v)
dv)\partial{'''}(\log\lambda(v) dv)}{d\log\lambda(v)} \,.
\end{eqnarray*}
Here $\partial\,,\partial'$ and $\partial''$ are arbitrary vector fields on the
Hurwitz space $H_{g, N}(k_1, \dots, k_l).$
\end{thm}

\medskip

\noindent The formula for $g$ appeared in \cite{dubrovin1}
while the formula for $c^\star$ follows immediately from the results in
\cite{dubrovin2}. Note that with the specific dependence of $u$ in the superpotentials
(\ref{Anpotential}) and (\ref{Bnpotential}) we have normalized $g$ and ${c}^\star$ so that $\partial_u$ is the
unity vector field (rather than $\frac{1}{2 \pi i} \partial_u$). Thus
${c}^\star\,(\partial_u,\partial{'},\partial{''})=g(\partial',\partial{''})\,.$
This also avoids a proliferation of $(2\pi i)$-factors in the final result.

Certain Hurwitz spaces are isomorphic to certain orbit spaces \cite{dubrovin1}.
For example,
\begin{eqnarray*}
\mathbb{C}^N/A_N &\cong& H_{0,N+1}(N+1)\,,\\
\mathbb{C}^{N+1}/{{\tilde A}^{(k)}_N} & \cong & H_{0,N+1}(k,N-k)\,,\\
\Omega/J(A_N) & \cong &H_{1,N+1}(N+1)\,.
\end{eqnarray*}
Thus the tower of generalizations mentioned in the introduction has a unified
description, at least for the $A_N$-cases, in terms of the theory of Hurwitz spaces.
This also leads to a way to expand the tower further via higher genus Hurwitz spaces, the
most natural being the space $H_{g,N+1}(N+1)\,.$
The $B_N$ examples come from introducing a $\mathbb{Z}_2$ grading onto the Huwitz space
(e.g. the superpotentials above have a $z\leftrightarrow -z$ symmetry).

\begin{example}\label{anbnexample}
{~~~}

\begin{itemize}
\item[(a)] Using the superpotential (\ref{Anpotential}) one obtains the intersection form and
(dual) prepotential for the $A_N$-Jacobi group orbit space in Theorem \ref{RileyStrachan} above \cite{RS}:

\begin{eqnarray*}
g & = & 2 du\,d\tau - \left.\sum_{i=0}^N (dz^i)^2\right|_{\sum_{j=0}^N z^j=0}\,\\
F^\star(u\,,{\bf z}\,,\tau) & = &
\frac{1}{2} \tau u^2 -
\frac{1}{2} u \left.\sum_{i=0}^N (z^i)^2\right|_{\sum_{j=0}^N z^j=0}
\\
& &
+\frac{1}{2}{\sum_{i\neq j}} f(z^i-z^j,\tau)-(N+1){\sum_i} f(z^i,\tau)
\end{eqnarray*}
where this function is evaluated on the plane
$\sum_{i=0}^N z^j=0\,.$
\medskip
\item[(b)] Using the superpotential (\ref{Bnpotential}) one obtains the intersection form
and
(dual) prepotential for the $B_N$-Jacobi group orbit space:
\begin{eqnarray*}
g & = & 2 du\,d\tau - 2\sum_{i=1}^N (dz^i)^2\,,\\
F^\star(u\,,{\bf z}\,,\tau) & = &
\frac{1}{2} \tau u^2 -
2 u \sum_{i=0}^N (z^i)^2\\
&&
+\sum_{i\neq j} \left\{f(z^i+z^j,\tau)+f(z^i-z^j,\tau) \right\}\\&&+ \sum_i f(2 z^i,\tau) - 2N \sum_i f(z_i,\tau)\,.
\end{eqnarray*}
\end{itemize}
\end{example}
Combining this with the earlier results on elliptic $\vee$-systems gives:
\begin{thm}
The elliptic $\vee$-systems given in sections (\ref{Ansolution})
and (\ref{Bnsolution}) define prepotentials that are the
almost-dual prepotentials associated to the $A_N$ and $B_N$ Jacobi
group orbit spaces.
\end{thm}
The form of this result, coupled with the examples of elliptic $\vee$-systems
leads to the following conjecture:
\begin{conjecture}
\label{conjecture}
Let $W$ be a Weyl group. For the Jacobi group orbit space
$
\Omega/J(W)
$
the dual prepotential takes the form (\ref{F}) with $h^\vee_\mathfrak{U}=0\,.$ Furthermore, $\mathfrak{U}=\mathcal{R}_W \cup \mathcal{R}_W^{irreg}\,$ (or its dual) and where
\[
\mathcal{R}^{irreg}_W=\{w(\Delta) | w \in W\} \quad {\rm or~~} \mathcal{R}^{irreg}_W=\{\pm w(\Delta) | w \in W\}
\]
for some weight vector $\Delta\,.$
\end{conjecture}

\noindent The conjecture seems plausible. It is true for the $A_N$ examples
and the $B_N$ examples (if one uses the dual root system) and {\em if} all orbit spaces are to behave in
the same generic way in the trigonometric limit then one must have
$h^\vee_\mathfrak{U}=0$ from the results of section
\ref{triglimits}.

One possible approach to proving this conjecture would be to show
that if a prepotential $F$ lies at the fixed point of the
involutive symmetry then so does the corresponding almost dual
prepotential. Since this is true for the Jacobi group examples
this would then prove the first part of the conjecture, but not
the second part on the structure of the set $\mathfrak{U}\,.$ The
Saito construction of Jacobi groups has recently been studied in
detail \cite{Satake}. Perhaps a formulation of a dual version of the result
would provide a proof of the conjecture.

Within the class of elliptic $\vee$-systems there remains the
problem of constructing examples with $h^\vee_\mathfrak{U}=0\,.$
For $W=G_2\,,F_4$ one may set $h=0\,,$ but for $E_{6,7,8}$ one
would have to append an $\mathcal{R}^{irreg}_W$ set of vectors. These
cases also remain problematical. If $h=0$ in the $G_2$ case one obtains a
dual prepotential that is actually the dual prepotential for the
$A_2$ Jacobi group orbit space, leaving a problem as to what the
correct $G_2$ solution would be. This case lies in the so-called
co-dimension one case and deserves closer study. The $G_2$ Jacobi
forms have also been constructed explicitly \cite{B} so it may be
possible to find the dual prepotential in this case by direct
calculation. It is also possible the the dual prepotential is the
same in these two cases: the reconstruction of the Frobenius
manifold from the dual picture requires additional data besides
the almost dual prepotential.

\section{Comments}

The idea of an elliptic $\vee$-system may clearly be studied further. As well as the obvious question on the relationship
between the functional ansatz and Jacobi group orbit spaces summarized in Conjecture \ref{conjecture}, there are many
other questions and problems that could be addressed. Given a complex Euclidean $\vee$-system one may study their restriction
to lower dimensions and the conditions required for the restricted system to also be a complex Euclidean $\vee$-system.
Clearly the same question can be asked for elliptic $\vee$-systems. Examples along this line may be obtained
from the restriction of the $A_N$ and $B_N$ Jacobi-group spaces to discriminants. This is achieved by introducing
multiplicities into the $A_N$ superpotential (\ref{Anpotential}),
\[
\lambda(p) = e^{2 \pi i u} \prod_{i=0}^m \left(\frac{\vartheta_1(v-z_i,\tau)}{\vartheta_1(v,\tau)}\right)^{k_i}\,,
\]
where $\sum_{i=0}^m k_i=N+1\,, \sum_{i=0}^m k_i z_i = 0\,,$ or on more general Hurwitz spaces $H_{1,N}(n_1\,, \ldots \,, n_m)$ and their
discriminants. Partial results have been obtained in \cite{thesis}, and these provide further examples of elliptic $\vee$-systems.
In fact, interesting examples of $\vee$-system can be found by looking on the induced structures on disciminants \cite{FV,iabs}
and clearly the same ideas could be applied here.

Possible applications of these solutions should come from Seiberg-Witten theory and the perturbative limits of such theories. This link is well known for rational
and trigonometric solutions, and the interpretation of the elliptic solutions found in \cite{RS} in terms of a 6-dimensional
field theory has been given in \cite{BMMM}, and one would expect similar results for the more general solutions constructed here (though
\cite{BMMM} does use the existence of a superpotential which is lacking for general solutions constructed here).

The tower of generalizations mentioned in the introduction clearly does not have to stop at elliptic solutions. An arbitrary Hurwitz
space $H_{G,N}(k_1\,,\ldots\,, k_l)$ carries the structure of a Frobenius manifold and hence an almost-dual structure. An interesting
question is whether or not there is an orbit space construction for these more general spaces:
\[
\begin{array}{ccccccccc}
H_{0,N}(N) & \longrightarrow & H_{0,N}(k,N-k) & \longrightarrow & H_{1,N}(N) & \longrightarrow & \ldots & \longrightarrow &H_{g,N}(N) \\
&&&&&&&&\\
\updownarrow&  & \updownarrow &  &  \updownarrow & & & & \updownarrow\\
&&&&&&&&\\
\mathbb{C}^N/A_N & \longrightarrow & \mathbb{C}^{N+1}/{\widetilde{A}_N^{(k)}} & \longrightarrow & \Omega/J({A_N})& \longrightarrow & \ldots & \longrightarrow &
\begin{array}{cc}
{\rm orbit}\\
{\rm space}\\
{\rm structure?}
\end{array}
 \end{array}
\]
It seems sensible to conjecture that such an orbit space exists. One would expect Siegel modular forms to play a role
instead of the modular forms used here. Higher genus Jacobi forms certain have been studied, but their use has yet to
percolate into the theory of integrable systems. The development, and applications of, the neo-classical $\vartheta$-function identities studied in Section \ref{FSconditions} remains to be done systematically. Certain higher genus analogues of these identities certainly exist,
since there exist almost-dual prepotentials on these Hurwitz spaces which, by construction, satisfy the WDVV equations.
In the genus 0 and genus 1 cases, the prepotential is very closely related to the prime form on the Riemann surface.
This may be the starting point for the development of a functional ansatz for the higher genus cases. Central to the
results presented here are the quasi-periodicity and modularity properties of the elliptic polylogarithm, and these
were obtained from the analytic properties of this function; the only role the analytic properties play were in the
development of these transformation properties. It would be attractive if one could obtain these directly from the
geometric properties of the prime form. This approach could then be used in the higher genus case where the analytic
properties are likely to be considerably more complicated.

Mention has been made already of the beautiful paper \cite{Etingof}. It would be interesting to see if
the ideas developed here could be used in the study of KZ and Dunkl-type systems. The idea would be to study
objects such as
\[
\sum_{\alpha,\beta \in \mathfrak{U}} \left[ f^{(3,0)}({\bf z}_\alpha) s_\alpha\,,f^{(3,0)}({\bf z}_\beta) s_\beta \right]
\]
where $s_\alpha$ and $s_\beta$ are shift operators. Conjecturally this quadratic term would be related to linear
terms in the function $f^{(2,1)}\,.$ The rational limit would then coincide with the classical work of Dunkl \cite{Dunkl}. Such
a development would be different to the elliptic Dunkl operators in the pioneering work of Buchstaber et al. \cite{Buchstaber}.
For a preliminary discussion of these ideas, see \cite{iabs2}\,.

Finally, one thing that has been learnt from this work is that on going from rational and trigonometric structure related to a Weyl group $W$ via
the root system $\mathcal{R}_W$ to elliptic structures, generalizations based entirely on the use of the root system $\mathcal{R}_W$ alone may not suffice.

\section*{Acknowledgments} I would like to thank Harry Braden, Misha Feigin and Andrew Riley for their comments on this paper.
I am also very grateful to the referee for pointing out certain errors in the original version, and for his/her careful reading of the
manuscript - this has resulted in much improved final version.


\end{document}